\documentclass[acmsmall,screen,nonacm]{acmart}

\usepackage{booktabs}
\usepackage{calc}
\usepackage{etoolbox}
\usepackage{longtable}
\usepackage{tabularx}
\usepackage{array}
\usepackage{makecell}
\usepackage{enumitem}
\usepackage{ragged2e}
\usepackage{xspace}
\usepackage{xurl}
\usepackage{pifont}

\newcolumntype{Y}{>{\raggedright\arraybackslash}X}
\newcolumntype{L}[1]{>{\raggedright\arraybackslash}p{#1}}

\newcommand{\cmark}{\ding{51}} 
\newcommand{\xmark}{\ding{55}} 

\graphicspath{{./figures/}}

\newcolumntype{Y}{>{\raggedright\arraybackslash}X}

\newcommand{\archlp}{Local-paged\xspace}
\newcommand{\archdp}{Disagg-pipeline\xspace}
\newcommand{\archss}{Shared-store\xspace}
\newcommand{\archmp}{Memory-pool\xspace}
\newcommand{\archht}{Hybrid-tier\xspace}

\AtBeginEnvironment{longtable}{%
  \scriptsize
  \setlength{\tabcolsep}{2pt}%
  \renewcommand{\arraystretch}{1.16}%
  \RaggedRight
}

\setcopyright{none}
\renewcommand\footnotetextcopyrightpermission[1]{}

\title{From Tensor Buffer to Distributed Memory Hierarchy: A Survey of KV Cache Management for LLM Serving}

\author{Jie Li}
\email{jie.li@ttu.edu}
\affiliation{%
  \institution{Texas Tech University}
  \department{Department of Computer Science}
  \city{Lubbock}
  \state{Texas}
  \country{USA}
}

\author{Tongyang Wang}
\email{tongyang.wang@ttu.edu}
\affiliation{%
  \institution{Texas Tech University}
  \department{Department of Computer Science}
  \city{Lubbock}
  \state{Texas}
  \country{USA}
}

\author{Yong Chen}
\email{yong.chen@ttu.edu}
\affiliation{%
  \institution{Texas Tech University}
  \department{Department of Computer Science}
  \city{Lubbock}
  \state{Texas}
  \country{USA}
}

\thanks{This research is supported in part by the National Science Foundation under grant OAC-2404438, and CNS-1939140 (A U.S. National Science Foundation Industry-University Cooperative Research Center on Cloud and Autonomic Computing).}



\begin{document}

\begin{abstract}
The key-value (KV) cache has become a first-order memory object in LLM serving rather than a temporary per-request tensor. This survey classifies more than thirty KV-management systems and frameworks using four axes: locality, lifetime, ownership, and substrate. The axes reveal five architectural archetypes---local-paged, disaggregated-pipeline, shared-store, memory-pool, and hybrid-tier. Once workload and hardware are fixed, ownership accounts for much of the remaining design variance among distributed systems. The survey also audits current evaluations and identifies seven missing KV-specific measurements, linking them to open problems in fault tolerance, isolation, tiered eviction, speculative decoding, MoE serving, and shared-cache semantics.

\end{abstract}

\maketitle

\section{Introduction}
\label{sec:introduction}

The key-value (KV) cache is a major dynamic memory consumer in autoregressive LLM inference~\cite{brown2020gpt3,hoffmann2022training,kwon2023vllm}, particularly for long-context and high-concurrency serving. For a decoder-only model with $L$ layers, $H_{\text{kv}}$ KV attention heads, head dimension $D$, sequence length $S$, and $b$ bytes per KV element, the aggregate per-request KV cache footprint is
$2 \times L \times H_{\text{kv}} \times D \times S \times b$ bytes, before accounting for tensor-parallel sharding, allocator fragmentation, metadata, or paging overheads. The factor of two accounts for separately stored keys and values. For a Llama-3.1-70B-style model with $L=80$, $H_{\text{kv}}=8$, $D=128$, and a 128K-token context, this corresponds to roughly 40~GiB per request in BF16/FP16. At 32K context, each request still requires about 10~GiB, so one hundred concurrent requests approach 1~TiB of aggregate KV state. As workloads shift toward long-context prompting, multi-turn interaction, and retrieval-augmented generation (RAG), the KV cache has become a first-order serving bottleneck rather than a small temporary buffer confined to the producing GPU.

This survey argues that the community is in the middle of an architectural shift. The KV cache was once a local tensor freed at request completion. Modern systems increasingly treat it as an explicit scheduling and memory-management object. Some systems remain local but virtualize, schedule, evict, or compress KV to raise effective capacity. Others move KV across workers, requests, memory tiers, or pooled-memory substrates, making it a distributed memory tier with explicit placement, lifetime, ownership, and substrate policies. These mechanisms are related responses to four questions: where the cache lives, how long it survives, who owns it, and what carries or exposes it. The survey's organizing contribution is to show that the classified systems, frameworks, and KV-reduction techniques---roughly three dozen entries enumerated in Tables~\ref{tab:all_systems_local}~and~\ref{tab:all_systems_dist}---concentrate around five recurring archetypes when sorted by their answers.

\subsection{Why the Local KV Abstraction Is No Longer Sufficient}

Three workload trends push KV state across boundaries that a per-GPU abstraction cannot cross. Context windows have grown from 1K--2K tokens in early GPT-style models to 4K in Llama~2~\cite{touvron2023llama2}, 128K in Llama~3.1~\cite{dubey2024llama3}, and longer in recent commercial systems; because KV memory scales linearly, a model needing 1.25~GiB of KV at 4K reaches 40~GiB at 128K and exceeds 300~GiB per request at 1M context. Chat and retrieval-augmented generation reuse long system prompts and retrieved contexts across requests, so a strictly per-request KV model recomputes the same prefix on every turn. Agentic workflows~\cite{wang2026forkkv} branch over intermediate states and resume earlier reasoning contexts, so KV state benefits from persisting beyond a single linear decode stream. KV state increasingly crosses at least one boundary~\cite{jin2024pdserve}---between GPUs, between nodes, between memory tiers, or between requests/sessions---and a serving system has to define the semantics and cost of each crossing.

\subsection{Six Design Responses}

Six lines of work emerged or accelerated during 2023--2025 in response to this redefinition: local KV virtualization and phase-aware scheduling, KV reduction by eviction or compression, prefill/decode disaggregation, cross-request prefix and context reuse, KV offload tiering, and pooled-memory substrates. We summarize each line and its anchor systems in detail in~\S\ref{sec:projections} rather than introducing the same enumeration twice. The six lines share four design decisions: where the cache lives, how long it must survive, who owns it, and what substrate carries or exposes it. Section~\ref{sec:taxonomy} develops these decisions into a taxonomy.

\subsection{Thesis}

Recent LLM serving systems no longer treat the KV cache only as an opaque per-request buffer. Local systems make KV explicit through paging, virtual memory, scheduling, eviction, and compression; distributed systems further expose KV through placement, ownership, lifetime, and substrate choices. These mechanisms form a small number of recurring architectures. End-to-end serving metrics were not built to show which architectural choice paid off; closing that gap takes a taxonomy, an archetype map, and a measurement-gap audit.

\subsection{Contributions}

The survey contributes a four-axis taxonomy of KV-cache management (locality, lifetime, ownership, substrate), with movement model, granularity, and scheduling objective treated as derived properties; an argument that the classified systems concentrate around five recurring architectural envelopes (\archlp, \archdp, \archss, \archmp, \archht), with within-archetype variance and boundary systems documented explicitly; a system-level classification distinguishing anchor systems, framework modes, and boundary cases; and a measurement-gap audit linking each open design question to the specific measurements needed to answer it.

\subsection{Survey Scope and Methodology}

The survey is selective but systematic. A system, framework, or technique is included when it materially changes at least one KV-management semantic---placement, lifetime, ownership, substrate, scheduling, reuse, or footprint---and is excluded when its main contribution is model-weight quantization, training-time KV behavior, generic scheduling without KV awareness, pure benchmark reporting, or attention approximation without a managed KV object. PagedAttention-style allocation, continuous batching, and KV compression are included when they define the baseline used by later distributed systems. Speculative decoding~\cite{leviathan2023fast,miao2023specinfer,cai2024medusa} is included because draft-token validation creates non-linear KV trajectories; MoE serving is included where expert-parallel placement or traffic constrains KV locality; multi-tenant isolation and fault tolerance are in scope because shared-store, offload, and hybrid-tier systems rely on remote or reusable KV state without specifying isolation or recovery semantics.

Candidate papers were drawn from arXiv, USENIX venues, MLSys, ASPLOS/ISCA/MICRO/HPCA, SOSP, SIGCOMM, SIGMETRICS, ACM DL, IEEE Xplore, Google Scholar, Semantic Scholar, and NeurIPS/ICML/ICLR systems tracks, with searches closed in May 2026 and keyword groups covering distributed KV serving, prefill/decode disaggregation, KV offload and tiering, CXL memory for LLM inference, prefix reuse, quantization, and continuous batching; snowballing from vLLM, DistServe, and Mooncake captured adjacent work. The classified population is summarized in Tables~\ref{tab:all_systems_local}~and~\ref{tab:all_systems_dist}, with adjacent work discussed in related work rather than classified. We use \emph{anchor system} for a system whose contribution defines an archetype; anchors are bolded in the system tables. Because the population mixes peer-reviewed systems, preprints, vendor reports, and production frameworks, the system tables include an evidence-tier column, and the prose hedges quantitative claims drawn from non-peer-reviewed sources.
\section{Background: KV Lifecycle and Serving Constraints}
\label{sec:background}

KV-cache management is shaped by three constraints: the autoregressive inference procedure that creates and consumes KV state, the memory hierarchy that stores or moves it, and the latency metrics used to evaluate serving systems. This section reviews the prefill/decode lifecycle (\S\ref{sec:prefilldecode}), the resulting memory and bandwidth scaling (\S\ref{sec:memorymath}), the block-table abstraction behind modern local KV managers (\S\ref{sec:pagedattention}), the parallelism layouts that restrict KV placement (\S\ref{sec:parallelism}), the main design responses in recent systems (\S\ref{sec:projections}), and the metric blind spots that motivate the audit in \S\ref{sec:gaps} (\S\ref{sec:metrics}).

\subsection{Transformer Inference: Prefill and Decode}
\label{sec:prefilldecode}

\begin{figure}[t]
 \centering
 \IfFileExists{figures/prefill-decode.pdf}{%
 \includegraphics[width=0.85\columnwidth]{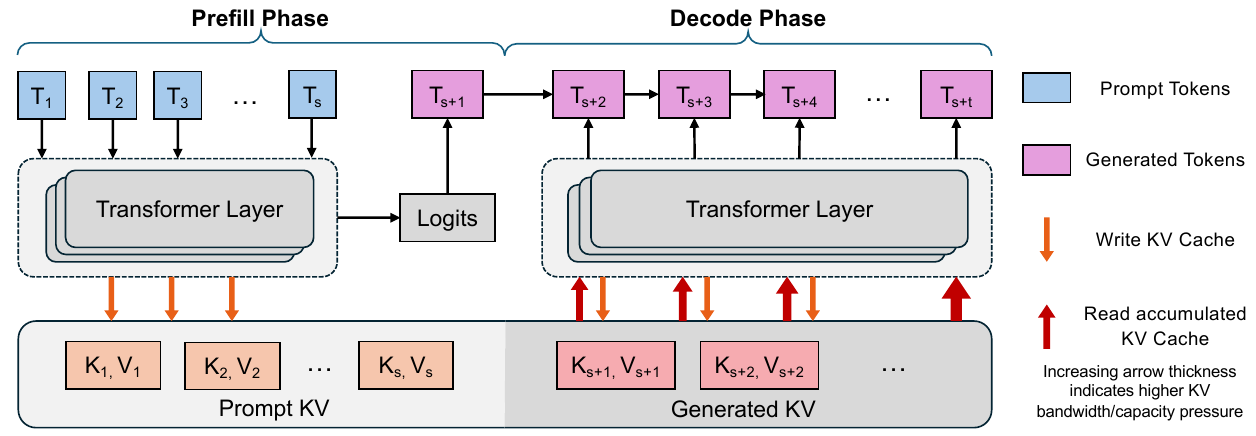}%
 }{%
 \fbox{\parbox{0.92\columnwidth}{\centering Prefill processes prompt tokens and writes KV; decode reads stored KV and appends one KV entry per generated token.}}%
 }
 \caption{Prefill materializes key and value vectors for prompt tokens; decode then advances one token per step through a shared Transformer stack, autoregressively feeding each newly emitted token back as the query for the next step. Each decode step reads the entire accumulated KV cache (prompt + previously generated entries) and appends one new $K_t, V_t$ pair; the cumulative KV-read traffic over $T$ decode steps therefore scales as $O(TS + T^2)$ (annotated in the figure), making decode increasingly sensitive to KV bandwidth and capacity as sequence length and concurrency grow.}
 \label{fig:phases}
\end{figure}

Transformer LLMs generate text autoregressively, one token at a time~\cite{vaswani2017attention}. Serving a request is usually decomposed into two phases (Figure~\ref{fig:phases}).

\textbf{Prefill.}
The model processes the input prompt in one or more forward passes. Under causal masking, each prompt token attends only to earlier prompt tokens, and each attention layer writes the token's key and value vectors into the KV cache. In dense attention, the attention work for a prompt of length $S$ is quadratic in $S$, although optimized kernels and distributed attention reduce constants and improve hardware utilization~\cite{dao2022flashattention,dao2024flashattention2,liu2024ringattention,tay2022efficient}. Prefill is therefore usually compute- and throughput-oriented: its efficiency depends on matrix-multiplication utilization, attention-kernel efficiency, batch size, prompt length, and the parallelism layout.

\textbf{Decode.}
After the first output token is selected, generation advances one token per step. At decode step $t$, the new query attends to all cached keys and values from the prompt and the $t$ previous generated tokens, then appends one new key/value entry per layer. Decode has lower arithmetic intensity than prefill because it performs small, step-wise computation while repeatedly streaming KV state from memory. It is often limited by KV-read bandwidth, KV capacity, scheduling overhead, and interconnect latency when KV state is remote or offloaded~\cite{liu2023blockwise}.

The two phases therefore prefer different operating points. Prefill benefits from large effective batches and high compute throughput; decode benefits from low-latency scheduling, high KV-read bandwidth, and enough memory to keep long-lived cache state close to the attention kernels. This asymmetry motivates phase-aware scheduling, prefill/decode disaggregation, and KV tiering, but the payoff depends on workload mix, interconnect cost, and system scale~\cite{kwon2023vllm,zheng2023sglang,zhong2024distserve,qin2024mooncake}.

\textbf{Capacity is linear; cumulative decode traffic is not.}
At any instant, the KV capacity required for one request is proportional to the number of resident tokens. For a prompt of length $S$ and $T$ generated tokens, the final cache contains $S+T$ token positions. Over the request lifetime, however, decode repeatedly scans the growing prefix. Across $T$ decode steps, the number of cached-token positions read is
\[
\sum_{t=0}^{T-1}(S+t) = TS + \frac{T(T-1)}{2},
\]
so cumulative KV-read traffic scales as $O(TS+T^2)$ under standard dense-attention decode.

For example, a 70B grouped-query-attention (GQA) model with roughly 320~KiB of BF16/FP16 KV state per token uses about 40~GiB of KV for a 128K-token prompt when ``K'' denotes 1024 tokens. If that request generates 32K tokens, dense decode reads on the order of
\[
320~\mathrm{KiB} \times \left(32\mathrm{K}\times128\mathrm{K} + \frac{32\mathrm{K}\times(32\mathrm{K}-1)}{2}\right),
\]
or about 1.4--1.6~PB of cumulative KV traffic, depending on binary versus decimal units and exact model configuration. Thus, KV bandwidth demand can be as important as KV capacity.

\subsection{KV Memory Footprint: Math, Scaling, and Tier Comparison}
\label{sec:memorymath}

\begin{figure}[t]
 \centering
 \includegraphics[width=0.85\columnwidth]{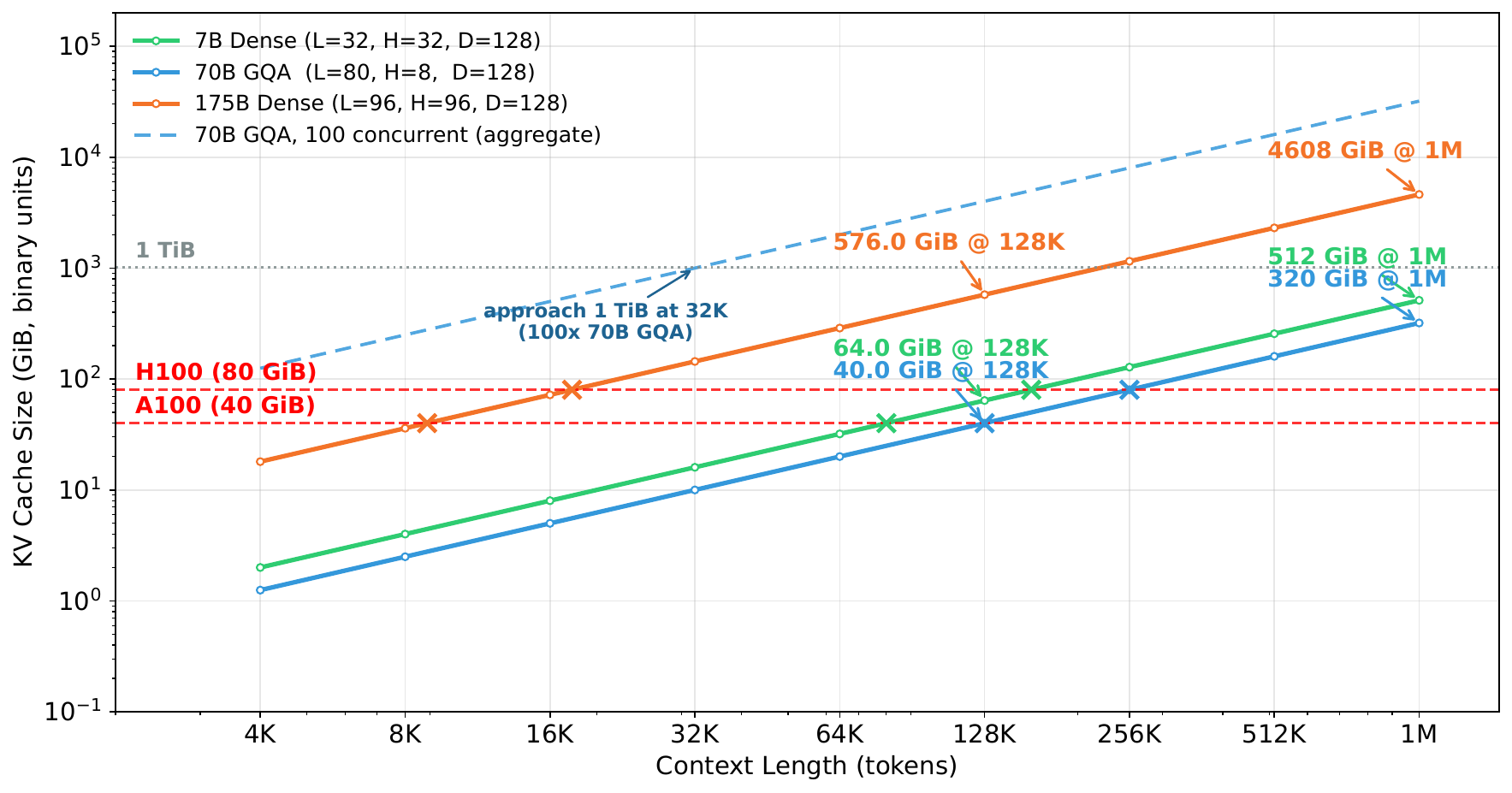}
 \caption{KV cache footprint versus context length for representative LLM architectures in FP16/BF16. Solid lines show per-request KV state for 7B Dense, 70B GQA, and 175B Dense; the dashed line shows aggregate KV state for one hundred concurrent 70B GQA requests, which approaches the 1~TiB dotted reference already at 32K context. Dashed red lines mark representative A100 (40~GiB) and H100 (80~GiB) HBM capacities, with crosses ($\times$) at each per-request HBM crossing. Sizes are in binary units (GiB); per-request capacity scales linearly in $S$, while cumulative decode traffic scales as $O(TS+T^2)$ (Figure~\ref{fig:phases}).}
 \label{fig:memory}
\end{figure}

For a decoder-only model with $L$ layers, $H_{\text{kv}}$ KV heads, head dimension $D$, resident sequence length $S$, and $b$ bytes per KV element, the aggregate per-request KV footprint is
\[
2 \times L \times H_{\text{kv}} \times D \times S \times b
\]
bytes. The factor of two accounts for separately stored keys and values. In FP16 or BF16, where $b=2$, this simplifies to $4 L H_{\text{kv}} D S$ bytes. This is the logical aggregate footprint before tensor-parallel sharding, allocator metadata, paging overheads, fragmentation, quantization, compression, or recomputation. Figure~\ref{fig:memory} shows the linear scaling across representative model families.

The same scaling applies across concurrent requests: aggregate KV memory grows with the sum of resident sequence lengths~\cite{chen2023posinterp,peng2023yarn,brandon2023striped}. A Llama-70B-style GQA model with $L=80$, $H_{\text{kv}}=8$, and $D=128$ needs roughly 40~GiB of aggregate KV state for one 128K-token request in BF16/FP16. At 32K context, it still needs about 10~GiB per request, so one hundred concurrent requests approach 1~TiB of aggregate KV state. This is far beyond what a single GPU can store together with model weights, activation scratch space, CUDA graphs, temporary buffers, and allocator slack.

\textbf{GQA reduces the footprint but does not remove the constraint.}
Grouped-query attention reduces $H_{\text{kv}}$ below the number of query heads, often to 8 KV heads in 70B-class models~\cite{ainslie2023gqa}. This proportionally reduces KV capacity and bandwidth relative to full multi-head attention. Yet long-context and high-concurrency workloads can still exhaust HBM, and prefix-caching or multi-turn serving may keep useful KV resident beyond one decode stream. Additional techniques such as layer-wise sharing, compression, eviction, or selective retention~\cite{wang2024squeezeattention,yang2024kvsharer} reduce the footprint further, but they also add representation, scheduling, and quality-control obligations.

\begin{table*}[t]
\centering
\caption{Memory tiers and interconnect paths available for KV placement or movement.
Values are representative; actual bandwidth and latency depend on device generation,
topology, NUMA placement, software path, and access granularity.
GPU HBM latencies are effective under kernel execution, not raw DRAM-array latencies.
NVLink/NVSwitch path costs in the sub-$\mu$s range apply to intra-node peer access;
cross-GPU access through NVSwitch on H100-class systems is typically several hundred ns and hop-dependent.
CXL latencies near the lower bound assume local-controller access;
rack-scale CXL access in TraCT-style deployments is closer to several hundred ns.
GPU access to host DRAM over PCIe incurs DMA setup cost in addition to transfer time.}
\label{tab:hierarchy}

\small
\setlength{\tabcolsep}{3.5pt}
\renewcommand{\arraystretch}{1.15}

\begin{tabularx}{\textwidth}{@{}L{2.8cm}L{1.8cm}L{2.8cm}L{3.2cm}Y@{}}
\toprule
\textbf{Tier / path} &
\textbf{Capacity} &
\textbf{Bandwidth} &
\textbf{Latency} &
\textbf{Access model} \\
\midrule

GPU HBM~\cite{nvidia2022h100,nvidia2023gracehopper,amd2023mi300} &
40--192~GB/GPU &
1--5~TB/s &
Hundreds of ns effective latency; access-pattern and kernel dependent &
Directly consumed by GPU kernels \\

\midrule

Peer GPU memory over NVLink/NVSwitch &
Node-scale GPU memory &
Hundreds of GB/s per GPU; TB/s aggregate &
Sub-$\mu$s to few-$\mu$s path cost; topology dependent &
GPU peer access, collectives, or explicit transfer \\

\midrule

RDMA-accessible remote GPU or host memory &
Cluster-scale &
Tens to hundreds of GB/s per endpoint &
Few-$\mu$s to tens of $\mu$s; software-path dependent &
One-sided or two-sided network transfer \\

\midrule

Host DRAM &
Hundreds of GB to TB/node &
Tens to hundreds of GB/s from CPU; lower effective GPU bandwidth over PCIe &
$\sim$100~ns CPU-local; higher from GPU due to PCIe, DMA, paging, or software overheads &
CPU load/store; GPU DMA, unified memory, or explicit copy \\

\midrule

CXL-attached memory~\cite{cxlconsortium2022} &
Device/rack-scale &
Tens to hundreds of GB/s &
Hundreds of ns to few $\mu$s; fabric dependent &
Load/store with fabric, NUMA, and coherence constraints \\

\midrule

SSD/NVMe &
TB--PB &
GB/s-scale &
Tens to hundreds of $\mu$s or higher &
Block I/O, file, object, or explicit staging interface \\

\bottomrule
\end{tabularx}
\end{table*}

Table~\ref{tab:hierarchy} summarizes the placement choices available to a KV manager. The tiers differ not only in capacity, bandwidth, and latency but also in access semantics: HBM is directly consumed by attention kernels, peer GPU memory requires topology-aware communication, host DRAM and SSD usually require staging, RDMA exposes remote memory through network operations, and CXL changes the load/store substrate while preserving contention, coherence, NUMA, and placement costs. These differences become the substrate dimension of the taxonomy in \S\ref{sec:taxonomy}.

\subsection{PagedAttention, Block Tables, and Continuous Batching}
\label{sec:pagedattention}

Before vLLM's PagedAttention, many serving systems allocated KV caches as contiguous tensors sized per sequence or for conservative maximum lengths~\cite{yu2022orca,pope2022efficiently}. Variable prompt and output lengths made this wasteful: memory reserved for unused future tokens could not be easily reused by other requests. PagedAttention introduced a block-based abstraction that later KV systems inherit~\cite{kwon2023vllm}.

PagedAttention allocates KV memory in fixed-size \emph{blocks}, often covering a small number of tokens, and maps logical token positions to physical blocks through a per-sequence \emph{block table}~\cite{kwon2023vllm}. Blocks can be allocated as a sequence grows and returned to a pool when no request, branch, or cached prefix needs them. This reduces external fragmentation and makes local KV memory schedulable.

Block indirection also enables sharing. In vLLM's original use case, parallel sampling and beam-search-style decoding can share prompt KV across continuations. In cross-request prefix caching, requests with identical prefixes can point to the same physical KV blocks and avoid redundant prefill work, subject to reference counting and copy-on-write or append-only semantics when sequences diverge.

SGLang's RadixAttention makes prefix identity explicit by storing reusable KV objects in a radix tree~\cite{zheng2023sglang}. A new request reuses KV along the longest matching prefix and computes only the unmatched suffix. This shifts KV management from anonymous tensor allocation to named, shared cache objects, introducing obligations around naming, reference counts, admission, eviction, and safe sharing. Those obligations reappear in the ownership axis of \S\ref{sec:taxonomy}.

Continuous batching is the corresponding scheduling mechanism~\cite{yu2022orca,kwon2023vllm}. Instead of waiting for all requests in a static batch to finish, the scheduler changes the active batch at decode-iteration granularity. This improves utilization but means requests with different prompt lengths, output lengths, priorities, and lifetimes share the same KV block pool. A block cannot be reclaimed while any active decode step, branch, or shared-prefix descendant depends on it. The interaction between continuous batching and distributed KV management is especially important for prefill/decode handoff, eviction safety, prefix-cache admission, and cache-aware routing.

\subsection{Parallelism Strategies and KV Placement Constraints}
\label{sec:parallelism}

KV placement is constrained by the model's parallelism layout. A system cannot freely choose single-GPU locality if attention is sharded across a tensor-parallel group, and it cannot freely centralize ownership if data-parallel replicas independently admit and schedule requests.

Tensor parallelism (TP) shards attention and feed-forward computations across GPUs; in common head-parallel implementations the KV cache is sharded by head across the TP group, and node-level TP over NVLink/NVSwitch is the typical layout because cross-node TP is collective-heavy~\cite{narayanan2021megatron,rasley2020deepspeed}. Pipeline parallelism (PP) partitions layers into stages, so KV is layer-local to the stage that owns the corresponding attention layers~\cite{narayanan2021megatron}. Data parallelism (DP) gives each replica an independent KV pool; without an external prefix cache or cache-aware router, repeated prefixes are recomputed across replicas~\cite{rajbhandari2020zero}. Expert parallelism (EP) shards feed-forward experts; in standard Transformer MoE designs KV still follows the attention-layer layout~\cite{fedus2022switch,jiang2024mixtral}, but expert all-to-all traffic contends with KV transfer for interconnect bandwidth, so EP affects KV management indirectly. Context- and sequence-parallel variants shard different dimensions and produce different KV-placement constraints.

Production systems often compose these strategies---TP within a node, PP or DP across nodes, EP for MoE---and long-context systems such as LoongServe and Infinite-LLM use sequence- or attention-oriented parallelism to reduce KV migration under variable context lengths~\cite{wu2024loongserve,lin2024infinitellm}. The taxonomy in \S\ref{sec:taxonomy} records KV choices alongside these parallelism assumptions, since the two are interact.

\subsection{Six Design Responses}
\label{sec:projections}

Recent systems respond to KV pressure through six recurring mechanisms. Two remain mostly within a local serving instance; the other four move KV management across boundaries.

P1, \emph{local KV virtualization and phase-aware scheduling}, includes vLLM's PagedAttention~\cite{kwon2023vllm}, vAttention's demand-backed virtual memory~\cite{pai2024vattention}, Orca's iteration-level scheduling~\cite{yu2022orca}, and SARATHI, Sarathi-Serve, DeepSpeed-FastGen, and TensorRT-LLM-style engines that mix prefills and decodes through chunking, split-fuse scheduling, in-flight batching, or paged KV pools~\cite{agrawal2023sarathi,agrawal2024sarathi,rasley2024fastgen}. The obligation is allocator--scheduler consistency.

P2, \emph{KV reduction by eviction, sparsification, quantization, or compression}, includes eviction and sparsification methods that retain selected tokens or allocate different budgets across layers and positions~\cite{zhang2023h2o,liu2023scissorhands,xiao2023streamingllm,li2024snapkv,cai2024pyramidkv}, quantization and compression methods that reduce bytes per retained element or compress semantically grouped context~\cite{liu2024kivi,hooper2024kvquant,kang2024gear,liu2025chunkkv}, and transfer-oriented compression such as CacheGen~\cite{liu2023cachegen}. The obligation is fidelity in generation quality and kernel compatibility.

P3, \emph{prefill/decode disaggregation}, places prefill and decode on separate workers connected by GPU peer links, RDMA, PCIe, CXL, or other substrates. DistServe, Splitwise, and KVDirect differ in placement and transfer-control choices~\cite{patel2023splitwise,zhong2024distserve,chen2025kvdirect}, and Trinity extends placement to include vector search alongside prefill and decode~\cite{liu2025trinity}. The obligation is KV handoff with controlled transfer, synchronization, admission, and ownership semantics.

P4, \emph{cross-request prefix and context reuse}, makes repeated system prompts, templates, few-shot examples, and RAG contexts cacheable. SGLang uses a radix tree, CacheBlend fuses cached knowledge for RAG~\cite{yao2024cacheblend}, LMCache adds content-based naming and storage abstractions~\cite{liu2025lmcache}, Preble uses prefix-aware scheduling~\cite{srivatsa2024preble}, and MemServe combines context caching with disaggregated inference and an elastic memory pool~\cite{hu2024memserve}. The obligation is safe identity and isolation: the system must decide when two requests may share or reuse KV~\cite{gim2024promptcache,juravsky2024hydragen} and prevent incorrect reuse across tenants, templates, model versions, adapters, or security domains.

P5, \emph{KV offload and tiering}, moves KV to host DRAM, SSD, remote memory, or storage when HBM is scarce. FlexGen is a historical precursor for offloading inference state across GPU, CPU, and disk~\cite{sheng2023flexgen}; Mooncake combines a disaggregated pipeline with a KVCache-centric store over CPU, DRAM, and SSD~\cite{qin2024mooncake}; ShadowKV keeps long-context KV in cheaper tiers while retrieving portions needed for sparse attention~\cite{sun2024shadowkv}; LMCache and SGLang HiCache-style modes belong here when KV persists across tiers. The obligation is migration control under deadline.

P6, \emph{memory pooling and disaggregated memory substrates}~\cite{li2022pond,sun2023demystifycxl,sharma2023cxlintro}, exposes capacity across workers through CXL-attached memory, RDMA-accessible pools, or specialized memory services. TraCT uses CXL shared memory as both a KV-transfer path and a rack-wide prefix-aware cache~\cite{yoon2025tract}; Beluga explores CXL-based KV-cache memory architecture~\cite{yang2025beluga}; CXL-SpecKV adds FPGA-side speculative prefetching over a CXL-disaggregated KV cache~\cite{liu2025cxlspeckv}. Pooling does not remove access cost: ownership, synchronization, consistency, contention, and failure handling remain runtime responsibilities.

Mooncake combines disaggregation, reuse, and tiering; SGLang combines virtualization and reuse and can extend toward tiering; CacheGen combines compression with transfer-oriented use cases; CXL-SpecKV combines reduction with pooled-memory substrates. Across combinations the same questions recur, which Section~\ref{sec:taxonomy} turns into the four taxonomy axes.

\subsection{Serving Metrics and Their Blind Spots}
\label{sec:metrics}

LLM serving papers commonly report latency and throughput metrics, but these rarely identify which KV-management mechanism caused an improvement. TTFT (time to first token) aggregates queueing, prefill, KV allocation, prefix-cache lookup, KV transfer when disaggregated, and decode-worker readiness, and does not reveal which component changed. TPOT (time per output token), and related inter-token-latency metrics, conflate local HBM bandwidth, remote-KV fetches, CPU/SSD prefetch misses, synchronization, scheduler delay, and offload recovery. Throughput alone ignores latency. Goodput---throughput of requests satisfying latency SLOs such as TTFT~$<t_1$ and TPOT~$<t_2$---joins the two, but it is highly workload- and SLO-dependent, and a goodput gain can come from admission control, larger batches, faster prefill, higher prefix-cache hit rate, faster KV transfer, fewer offload misses, or better decode scheduling. SLO attainment summarizes the fraction of requests meeting their targets but hides distributional behavior, and P50, P95, and P99 may be dominated by different sources of delay. Section~\ref{sec:gaps} identifies the additional measurements needed to make those mechanisms visible.
\section{Four-Axis Taxonomy}
\label{sec:taxonomy}

Four design choices recur across the systems surveyed: \emph{locality} (where a consumer obtains a KV block), \emph{lifetime} (how long the system attempts to keep the block useful), \emph{ownership} (which component allocates, names, evicts, and resolves contention for the block), and \emph{substrate} (which data-movement or memory-access mechanism carries or exposes the bits). These are the four axes of the taxonomy.

The taxonomy operates at the granularity of a KV \emph{block} when a system exposes block-based KV management---the PagedAttention allocation unit, often 16--32 tokens~\cite{kwon2023vllm}---and otherwise at the smallest KV unit the system places, transfers, names, or evicts. A system's design is described by an $(A, B, C, D)$ tuple, with the understanding that a system may occupy multiple positions on an axis (for example, A1 direct handoff for newly produced KV and A2 shared-store lookup for reusable prefixes).

\textbf{Classification rule.} We classify the dominant KV-management contribution of each system, framework mode, or technique. The procedure is operational: identify the KV object the system manages; determine whether the main contribution changes placement, lifetime, ownership, substrate, representation, or scheduling dependency; assign the primary tuple from that contribution; then add secondary labels only when the paper or implementation exposes a distinct mode, component, or boundary mechanism with different axis values. A system is marked as boundary when it changes KV movement or representation without defining a full ownership and lifetime contract. Thus, SGLang appears as local-paged in its core radix-tree serving mode and shared-store-adjacent in hierarchical-cache modes; CacheGen is boundary because it changes the representation and transfer path without defining a persistent shared namespace; and Mooncake is hybrid-tier because the full system composes disaggregation, shared storage, and tiering rather than belonging to only one archetype.

\subsection{Axis A: Locality}

Locality describes \emph{where} the consumer of a KV block obtains it. Table~\ref{tab:locality} defines the four locality levels.

\begin{table}[t]
\centering
\caption{Axis A: locality levels.}
\label{tab:locality}

\small
\setlength{\tabcolsep}{3pt}
\renewcommand{\arraystretch}{1.12}

\begin{tabularx}{\columnwidth}{@{}L{0.7cm}L{2.5cm}Y@{}}
\toprule
\textbf{Level} &
\textbf{Name} &
\textbf{Description} \\
\midrule

A0 &
Local-only &
KV is consumed on the GPU or worker that produced it or currently owns it. \\

A1 &
Direct remote &
KV is transferred directly from a producer or owner to a consumer, typically along one logical producer--consumer path. \\

A2 &
Shared store &
KV resides in a shared namespace or distributed store from which eligible workers can fetch blocks. \\

A3 &
Shared memory &
KV is accessed through a load/store-addressable memory region shared across workers; coherence, consistency, and synchronization are supplied by the fabric, software, or both. \\

\bottomrule
\end{tabularx}
\end{table}

Of the four axes, locality has the most direct effect on data-access latency and bandwidth pressure. A0 keeps remote data movement off the KV critical path. A1 adds a direct producer-to-consumer transfer. A2 adds lookup and store-access overhead, and A3 exposes a shared-memory access path whose behavior depends on the fabric, the placement policy, and the synchronization discipline. Aggregate capacity usually grows as a design moves from local-only KV toward shared-store or shared-memory designs, though the exact ordering depends on the deployment: a large A2 store can expose more aggregate capacity than a local design, while an A3 design is bounded by the size and topology of its shared pool.

Locality does not map one-to-one onto hardware. A1 is commonly built on RDMA or GPU-aware communication, but TCP or host-mediated copies serve less latency-sensitive paths equally well. A3 today is associated with CXL-like fabrics, yet the level is defined by shared load/store-addressability, so a future non-CXL fabric with the same access model would also qualify as A3.

\subsection{Axis B: Lifetime}

Lifetime describes \emph{the semantic scope over which a system attempts to keep a KV block useful}. Table~\ref{tab:lifetime} separates reuse scope from retention strength: B0-B2 give the intended reuse scope, and B3 and B4 are modifiers for whether retention is opportunistic or recoverable.

\begin{table}[t]
\centering
\caption{Axis B: lifetime levels.}
\label{tab:lifetime}

\small
\setlength{\tabcolsep}{3pt}
\renewcommand{\arraystretch}{1.12}

\begin{tabularx}{\columnwidth}{@{}L{0.7cm}L{2.5cm}Y@{}}
\toprule
\textbf{Level} &
\textbf{Name} &
\textbf{Description} \\
\midrule

B0 &
Per-request &
KV is needed only for the active request and can be discarded when the request finishes. \\

B1 &
Per-session &
Reusable KV is retained across multiple turns of one user session or conversation. \\

B2 &
Cross-session &
Reusable KV is retained across requests, sessions, users, or workloads that share prefixes or retrieved context. \\

B3 &
Opportunistic &
KV is retained while capacity and policy allow, without a hard reuse guarantee; this modifier can coexist with B0--B2. \\

B4 &
Durable/recoverable &
KV is backed by a documented recovery contract and can survive relevant node or process failures; this modifier can coexist with B0--B2. \\

\bottomrule
\end{tabularx}
\end{table}

Lifetime governs how long KV state stays resident, so it drives aggregate memory pressure more than any other axis. A B0 system allocates KV for the active request and releases it at completion. B1 retains reusable KV across the turns of one session, raising residency but avoiding repeated within-session prefill. B2 retains KV across sessions or users, amortizing prefill for shared prefixes and RAG contexts. B3 marks opportunistic retention: blocks are kept while capacity allows and dropped under pressure. Because B3 is a policy rather than a scope, it composes with B0–B2; an entry like B2+B3 means cross-session reuse held opportunistically.

No system in our population occupies B1 as a primary label. Two mechanisms are sometimes mistaken for it, but do not meet the definition. StreamingLLM-style attention sinks bound the KV footprint inside a single decode stream by keeping a few initial tokens plus a sliding window and discarding the middle, which is request-scoped footprint reduction (B0). SGLang's local RadixAttention retains finished-request KV in an LRU radix tree for prefix-keyed reuse across requests, which is B2 under opportunistic retention (B3). Deployed chat services do hold per-session KV across turns, but none in our population formalizes a per-session contract---a retention budget, cross-session eviction priority, and session isolation---separate from these two mechanisms. We keep B1 in the axis as a real design point and record its absence as DG7 (~\S\ref{sec:agenda}). B4 is likewise empty: no surveyed system provides a documented, recoverable KV-state contract across node or store failure, a gap ~\S\ref{sec:agenda} takes up.

\subsection{Axis C: Ownership}

Ownership describes \emph{who} allocates, evicts, names, and arbitrates access to a KV block. Table~\ref{tab:ownership} lists the ownership levels. Hardware does not settle this question — the same substrate can be governed by any of several control planes---which is why ownership is where engineering trade-offs surface most clearly. C0 keeps control local to each worker, minimizing control-plane overhead at the cost of cross-worker optimization. C1 puts one logical component in charge: global policy becomes simple, but the component is a scalability and availability concern unless it is replicated or sharded. C2 distributes decisions across workers or cache nodes through a directory, hash mapping, lease, or peer-to-peer protocol, removing the single bottleneck while adding coordination complexity. C3 leaves ownership to software conventions over shared memory, which simplifies the data path but makes allocation, write discipline, and failure behavior harder to verify under concurrency.

The practical consequence is that two systems matched on locality, lifetime, and substrate can still part ways on who controls placement and eviction---and that choice, more than the others, sets their scalability and fault-tolerance profiles.

\begin{table}[t]
\centering
\caption{Axis C: ownership levels.}
\label{tab:ownership}

\small
\setlength{\tabcolsep}{3pt}
\renewcommand{\arraystretch}{1.12}

\begin{tabularx}{\columnwidth}{@{}L{0.7cm}L{2.5cm}Y@{}}
\toprule
\textbf{Level} &
\textbf{Name} &
\textbf{Description} \\
\midrule

C0 &
Per-worker &
Each worker manages its own KV blocks, with no global authority for placement or eviction. \\

C1 &
Coordinator &
A centralized control component decides KV placement, admission, eviction, or routing. \\

C2 &
Distributed &
Workers or cache nodes cooperate through a distributed directory, hash mapping, lease protocol, or peer-to-peer control protocol. \\

C3 &
Shared memory &
KV resides in a shared memory region, while software conventions define allocation, ownership, and write discipline. \\

\bottomrule
\end{tabularx}
\end{table}

\subsection{Axis D: Substrate}

Substrate describes \emph{how the bits move or are accessed}. A system can use multiple substrates simultaneously. Table~\ref{tab:substrate} compares the substrate labels used in the system tables. We use D-local for ordinary GPU HBM/local device memory and reserve D0--D6 for non-local transfer paths, pooled-memory paths, lower-bandwidth capacity tiers, or programmable data-path components that become explicit design choices.

\begin{table*}[t]
\centering
\caption{Axis D: substrate comparison.}
\label{tab:substrate}

\small
\setlength{\tabcolsep}{4pt}
\renewcommand{\arraystretch}{1.12}

\begin{tabularx}{\textwidth}{@{}L{1cm}L{3.2cm}YY@{}}
\toprule
\textbf{Level} &
\textbf{Name} &
\textbf{Access Model} &
\textbf{Typical Role} \\
\midrule

D-local &
GPU HBM/local memory &
Direct access by local GPU kernels &
Baseline residence for active KV \\

D0 &
NVLink/NVSwitch &
GPU--GPU copy, peer access, or collective communication within an NVLink/NVSwitch domain &
Intra-node or tightly coupled multi-GPU KV movement \\

D1 &
RDMA/GPUDirect RDMA over IB or RoCE &
One-sided or two-sided network transfer, often GPU-aware &
Low-latency cross-node KV transfer \\

D2 &
Transfer libraries or abstraction layers, e.g., NIXL or UCX &
Software abstraction over RDMA, TCP, shared memory, GPU-copy paths, or storage backends &
Portable KV-transfer layer rather than a separate physical substrate \\

D3 &
CXL-like shared memory &
Load/store access to memory attached through a coherent, partially coherent, or software-managed fabric &
Shared-memory KV pool or memory expansion \\

D4 &
Host DRAM over PCIe &
Pinned-memory access, CPU-mediated staging, or DMA copy between GPU and host memory &
KV offload or staging tier \\

D5 &
Local/remote SSD or storage system &
Block, file, object, or storage-fabric I/O &
Capacity tier for cold or reusable KV state \\

D6 &
SmartNIC/DPU/FPGA &
Programmable data movement, near-data processing, or in-network/data-path processing &
Offloaded transfer, indexing, compression, prefetching, or scheduling support \\

\bottomrule
\end{tabularx}
\end{table*}

Substrate is the axis most tightly bound to hardware availability. RDMA and GPUDirect RDMA dominate low-latency cross-node KV movement in disaggregated systems, and CXL-like fabrics are the natural fit for load/store-addressable shared memory. Transfer libraries such as NIXL and UCX sit above one or more of these substrates rather than constituting substrates of their own---vLLM, for instance, describes its NIXL connector as an asynchronous KV-transfer path for disaggregated prefilling~\cite{nixl2025dynamo}, and Mooncake's transfer engine moves KV and pools it across tiers, with performance set by the underlying transport~\cite{qin2024mooncake}. We group SmartNICs, DPUs, and FPGAs under D6 because all three expose programmable data-path or near-data compute to KV management, where their differences in bandwidth, programmability, or control-plane assumptions matter. Table~\ref{tab:hierarchy} gives representative capacity, bandwidth, and latency ranges for the major tiers, and these constraints feed directly into the archetype concentration in ~\S\ref{sec:archetypes}.

\subsection{Derived Properties}

The four axes also determine several design properties that are often used as classifiers in prior work. We treat these properties as \emph{derived} rather than independent because their values follow from combinations of locality, lifetime, ownership, and substrate.

The \emph{movement model} follows from locality and substrate: A1+D1 naturally maps to RDMA push or pull, A2 to lookup plus remote fetch, and A3+D3 to shared-memory access. \emph{Granularity} follows mainly from lifetime and ownership: B0+C0 systems can use request-local tensors or page blocks, while B2+C2 systems need reusable names for prefixes, blocks, or segments. The \emph{scheduling objective} follows from the full tuple: C1 disaggregated pipelines optimize global placement and goodput, C2 shared-store designs balance cache-hit rate against lookup and transfer cost, and C0 local-paged designs focus on batching, admission, and local memory pressure.

\subsection{Why These Four Axes}

The four axes satisfy an orthogonality test: a system can change one axis while leaving the others largely unchanged. DistServe and KVDirect illustrate the point: both address disaggregated inference and KV transfer over RDMA-like substrates, but they differ in how much control is centralized versus embedded in a distributed communication protocol~\cite{zhong2024distserve,chen2025kvdirect}. Two axis values---A3 (shared-memory locality) and C3 (shared-memory ownership discipline)---are usually induced when D3 (CXL-like shared memory) is the substrate. We retain them as separate axis values rather than folding them into D3 because they encode different consequences: A3 describes the data path that a consumer follows to reach a KV block, C3 describes the software discipline used to allocate and arbitrate that block, and D3 describes the fabric that carries it. A future system could plausibly expose A3+C2 (shared memory with directory-managed ownership) or A2+C3 (object-store interface over shared memory), and treating the three properties separately keeps the taxonomy expressive enough to record such designs.

A fifth candidate axis---\emph{granularity} (token, block, layer, tensor, or prefix)---is excluded because it is largely induced by lifetime and ownership. Similarly, \emph{consistency model} is excluded because most surveyed systems do not state an explicit KV consistency contract. Prior work on attention approximation, long-context attention, and efficient inference~\cite{shazeer2019fast,gu2022efficiently,jiang2024minference,ding2023longnet} is related to reducing KV pressure or attention cost, but orthogonal to the placement semantics considered here. Consistency may become a useful axis once systems document contracts for concurrent access, reuse, failure, and recovery.

\subsection{Comparison with Alternative Taxonomies}
\label{sec:taxonomy_alternatives}

Prior surveys typically organize this literature by optimization level, mechanism, or workload. Level taxonomies separate token-level, model-level, and system-level techniques but cannot distinguish system-level designs that differ in control-plane ownership. Mechanism taxonomies classify by substrate (RDMA, CXL, SSD, GPU-interconnect) cleanly but conflate systems whose lifetime and ownership differ; a substrate-only view makes LMCache-like shared caches and Mooncake-like KV-cache stores look closer than their control planes are. Workload taxonomies explain why a deployment cares about TTFT, TPOT, throughput, or reuse without identifying the architectural decision that produced the result. The four-axis tuple treats locality, lifetime, ownership, and substrate as separable, leaving sparse, linear, and long-context attention~\cite{kitaev2020reformer,wang2020linformer,zaheer2020bigbird,beltagy2020longformer} as complementary KV-workload techniques rather than primary classifiers.

\subsection{Worked Example: Classifying a System on the Four Axes}
\label{sec:worked_example}

The classification procedure is most easily seen on a concrete system. Infinite-LLM~\cite{lin2024infinitellm} is a recent distributed serving system for ultra-long contexts; it splits attention across workers using a context-parallel scheme and routes KV through a cluster-level memory pool rather than keeping it on the producing worker. Applying the procedure in~\S\ref{sec:taxonomy} step by step:

\begin{enumerate}
\item \emph{Identify the KV object.} The managed unit is the per-worker KV slice produced during distributed attention.
\item \emph{Determine the dominant contribution.} Infinite-LLM's primary contribution is making KV addressable across the cluster so that long contexts spill out of a single worker; the change is to locality and ownership rather than to substrate or lifetime semantics.
\item \emph{Assign the primary tuple.} Locality is A2 (shared store across workers, exposed as a logical KV pool rather than as a single shared-memory region), lifetime is B0 (request-scoped, because the long-context KV is consumed only by the originating request), ownership is C2 (distributed across the worker pool through a directory and routing protocol), and substrate is the underlying network plus host or device memory tiers, recorded as D-local+D1.
\item \emph{Add secondary labels only when warranted.} Infinite-LLM does not expose a separate cross-session reuse mode, so no secondary B label is added.
\item \emph{Mark boundary status.} The system is close to the shared-store envelope in (A, C) and to the disaggregated-pipeline envelope in (B, D); we therefore record it as a boundary case adjacent to shared-store and discuss it in~\S\ref{sec:archetypes} together with the boundary table.
\end{enumerate}

The same procedure applies to other systems and is the basis for the per-system rows in Tables~\ref{tab:all_systems_local}~and~\ref{tab:all_systems_dist}.
\section{System Archetypes}
\label{sec:archetypes}

The taxonomy in~\S\ref{sec:taxonomy} supplies the axes; this section uses them to make the concentration claim. The systems in Tables~\ref{tab:all_systems_local}~and~\ref{tab:all_systems_dist} concentrate around five recurring envelopes rather than spreading uniformly across the four-axis design space. The claim is empirical and qualitative: the archetypes are envelopes induced by repeated axis signatures across the classified systems, not a statistical clustering result. An \emph{anchor system} is a system whose contribution makes an archetype recognizable; a \emph{boundary system} spans archetypes or contributes a mechanism reused by several archetypes. The archetype names are architectural, not mechanism level: \archlp covers paging, virtual-memory, scheduling, eviction, and compression alike, because all of them keep ownership local while changing the local KV cost model. Two things follow. Mechanism diversity within an archetype is common, and ~\S\ref{sec:variance} quantifies it. And the \archht envelope is wider than the rest, since its members compose two or more mechanisms by design; ~\S\ref{sec:fivearch} tightens that definition.

\subsection{Why the Population Concentrates: Parallel Emergence}

The classified systems span several years and design-response families, yet their design points cluster more tightly than that diversity suggests. Hardware narrows practical choices---RDMA or GPU-aware networking dominates explicit cross-node KV movement, and CXL-like fabrics dominate load/store-addressable memory pools. Workload structure creates prefix and segment reuse from chat sessions, system prompts, and RAG pipelines, and makes prefill-to-decode handoff one-directional, large, and latency-critical. Serving-stack compatibility favors PagedAttention-style block granularity because block tables already support allocation, eviction, transfer, and reuse. Systems with different stated goals therefore converge on similar primitives because they face the same KV movement, naming, and scheduling constraints.

\subsection{The Five Archetypes}
\label{sec:fivearch}

\begin{figure*}[t]
 \centering
 \includegraphics[width=0.85\textwidth]{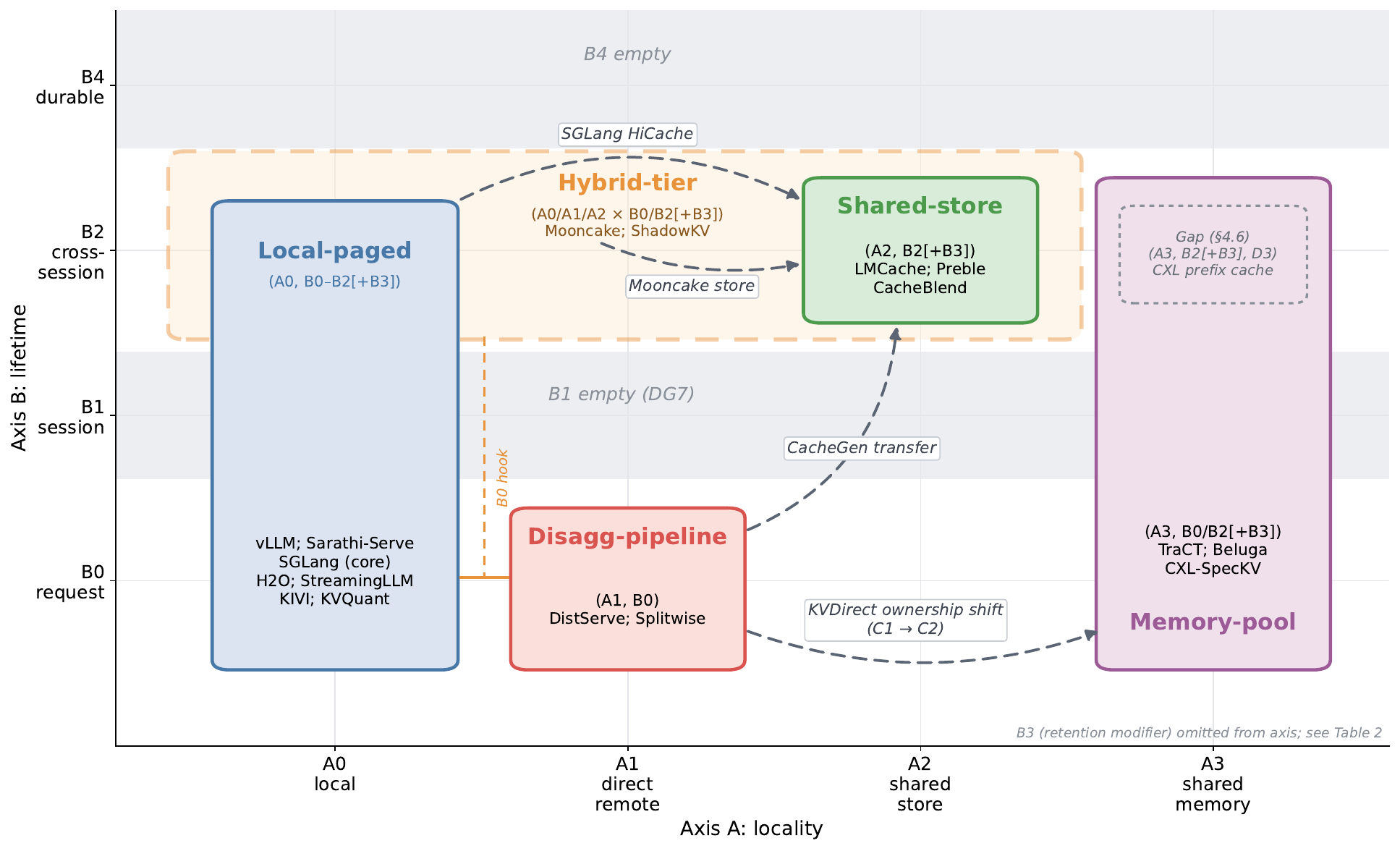}
 \caption{Archetype map in the $(A,B)$ projection. Colored regions give a visual map of the five archetype envelopes from Table~\ref{tab:archetypes}: Local-paged spans B0--B2 (per the table signature B0--B2[+B3]); Disagg-pipeline sits at B0; Shared-store at B2; Memory-pool at A3 spanning B0--B2; Hybrid-tier composes A0/A1/A2 with B0/B2, rendered as a dashed envelope sitting above the empty B1 band with a thin ``B0 hook'' line dropping into the request-scoped region (so the box is not visually anchored on B1). B3 is a retention modifier and is omitted from the vertical axis; Table~\ref{tab:archetypes} records where it applies. The B1 (per-session) and B4 (durable) rows are marked empty per~\S\ref{sec:taxonomy}. Dashed arrows mark labeled boundary/compositional paths (SGLang HiCache, Mooncake store, CacheGen transfer, KVDirect ownership shift). The (A3, B2[+B3], D3) CXL prefix-cache gap from~\S\ref{sec:crossarchetype} is marked inside the Memory-pool envelope.}
 \label{fig:scatter}
\end{figure*}

\begin{table}[h]
\small
\centering
\caption{The five archetypes with their $(A, B, C, D)$ signatures and anchor systems.}
\label{tab:archetypes}
\begin{tabular}{@{}p{2.5cm}p{7.5cm}p{3cm}@{}}
\toprule
\textbf{Archetype} & \textbf{Signature} & \textbf{Anchors} \\
\midrule
Local-paged & (A0, B0--B2[+B3], C0, D-local[+D0]) & vLLM, Sarathi-Serve \\
Disagg-pipeline & (A1, B0, C1/C2, D-local+D1) & DistServe, Splitwise \\
Shared-store & (A2, B2[+B3], C1/C2, D-local+[D1/D4/D5]) & LMCache, Preble \\
Memory-pool & (A3, B0/B2[+B3], C3, D-local+D3[+D6]) & TraCT \\
Hybrid-tier & (A0/A1/A2, B0/B2[+B3], C0/C1, D-local+[D1/D4/D5]) & Mooncake, ShadowKV \\
\bottomrule
\end{tabular}
\end{table}

The signatures in Table~\ref{tab:archetypes} are envelopes rather than exact tuples for every member; \S\ref{sec:systemtable} gives the per-system assignments. Figure~\ref{fig:scatter} is intentionally less precise than the table: it visualizes where the archetype envelopes sit in the $(A,B)$ projection, highlights the empty B4 region, and marks boundary or compositional paths among the regions.

\textbf{Archetype 1: Local-paged.} KV is managed within a single GPU, node, or local serving instance using block-table allocation, virtual memory, local scheduling, or local footprint reduction. There is no cross-node KV-store substrate and no distributed KV ownership in the dominant contribution. The anchors are vLLM~\cite{kwon2023vllm} (PagedAttention and continuous batching) and Sarathi-Serve~\cite{agrawal2024sarathi} (chunked prefill and stall-free scheduling). Orca, SARATHI~\cite{agrawal2023sarathi}, DeepSpeed-FastGen~\cite{rasley2024fastgen}, TensorRT-LLM-style engines, and vAttention~\cite{pai2024vattention} keep KV local while changing batching, preemption, virtual-memory backing, or prefill chunking; framework engines may also use intra-node GPU interconnects for broader model execution without becoming shared KV stores. H2O, Scissorhands, StreamingLLM, SnapKV, PyramidKV, KIVI, KVQuant, and ChunkKV reduce the local footprint through eviction, sparsification, quantization, or compression~\cite{zhang2023h2o,liu2023scissorhands,xiao2023streamingllm,li2024snapkv,cai2024pyramidkv,liu2024kivi,hooper2024kvquant,liu2025chunkkv}. SGLang~\cite{zheng2023sglang} is local-paged in its core serving mode, where prefix reuse remains within the serving instance, while FlexGen~\cite{sheng2023flexgen} is a local-control offload precursor whose GPU/CPU/disk tiering we group with \archht in the timeline (Figure~\ref{fig:timeline}). The persistence of this archetype shows that many deployments still benefit most from improving the local allocator, scheduler, or representation.

\textbf{Archetype 2: Disaggregated-pipeline.} KV crosses from specialized prefill workers to decode workers and normally remains request-scoped. DistServe~\cite{zhong2024distserve} and Splitwise~\cite{patel2023splitwise} are the anchors; adjacent systems include P/D-Serve~\cite{jin2024pdserve}, KVDirect~\cite{chen2025kvdirect} when its main role is transfer protocol design, LoongServe when elastic sequence parallelism reduces KV migration under long contexts~\cite{wu2024loongserve}, and hardware specialization such as SPAD~\cite{zhang2025spad}. This is the simplest distributed archetype because it introduces one primary locality crossing while avoiding a persistent global KV namespace.

The key engineering tension in this archetype is the granularity of the KV handoff. A system can transfer KV only after prefill completes, or it can pipeline KV movement as earlier layers or chunks become available. Pipelined handoff can reduce decode-worker idle time, but it increases message count, synchronization, and metadata-path overhead. Existing disaggregated-serving papers usually report end-to-end latency, goodput, or SLO attainment, but they rarely isolate the metadata and message-granularity costs of the handoff; this is a concrete instance of the metadata-gap audit in~\S\ref{sec:gaps}.

\textbf{Archetype 3: Shared-store.} KV resides in a global or semi-global namespace and can be reused across requests. LMCache~\cite{liu2025lmcache} and Preble~\cite{srivatsa2024preble} are the anchors. The key distinction within the archetype is ownership: some designs use a centralized coordinator, scheduler, or cache server, while others distribute lookup, placement, or transfer responsibilities. Adjacent systems include SGLang HiCache, CacheBlend~\cite{yao2024cacheblend}, MemServe, and CacheGen~\cite{liu2023cachegen,hu2024memserve} when their mechanisms support transfer or reuse but do not by themselves constitute a long-lived shared KV store. CacheBlend is shared-store-adjacent because it makes cached non-prefix RAG KV usable through selective recomputation rather than by defining a standalone store. Shared-store systems are diverse because they must name, locate, evict, and define safe reuse semantics for KV objects that outlive any single request.

Shared-store systems face a fundamental tension between cache hit rate and metadata overhead. Longer retention increases the probability that an incoming request can reuse a stored prefix or segment, but it also increases lookup cost, metadata memory, block-location tracking, and eviction complexity. Few shared-store papers report the metadata-to-data memory ratio for realistic workloads; this is MG2 in~\S\ref{sec:gaps}. The tension becomes acute at scale: a store with 100~TB of KV state and 1~TB of metadata is qualitatively different from a store with 1~TB of KV and 10~GB of metadata, and published evaluations rarely span this range.

\textbf{Archetype 4: Memory-pool.} KV is placed in a physically shared memory pool, typically CXL-attached memory, and workers access it through load/store, DMA, or memory-mapped operations. TraCT~\cite{yoon2025tract} is the anchor; Beluga~\cite{yang2025beluga}, CXL-SpecKV~\cite{liu2025cxlspeckv}, and the elastic memory-pool component of MemServe~\cite{hu2024memserve} are adjacent memory-pool designs. This archetype is small because CXL deployment is still limited, but it is architecturally distinct: the cost model shifts from explicit network transfer and object lookup to shared-memory placement, access latency, bandwidth contention, and software-managed coordination.

The key open question for this archetype is whether shared-memory coordination can be maintained at CXL scale without assuming transparent hardware coherence across the entire pool. A practical CXL KV system must still define ownership, allocation, invalidation, and reclamation semantics. TraCT-style designs show that rack-scale shared-memory KV paths are plausible, but no published system has yet established the scalability limits of software-managed consistency and allocation for cluster-scale KV pools.

\textbf{Archetype 5: Hybrid-tier.} A hybrid-tier system composes two or more of the previous archetypes' mechanisms under an integrated control plane that arbitrates among them. Mooncake~\cite{qin2024mooncake}, the canonical anchor, combines prefill/decode disaggregation, shared KV storage, and tiering across GPU, CPU, and storage media under a centralized control plane. We therefore restrict the hybrid-tier label to systems whose distinguishing contribution is the integration logic itself rather than the underlying components.

ShadowKV~\cite{sun2024shadowkv} sits at the edge of this definition: it combines local-control offload tiering with sparse retrieval, but its dominant path remains local-control offload rather than Mooncake-style distributed ownership. We record ShadowKV under hybrid-tier because it integrates two mechanisms (offload tiering and sparse retrieval) under a unified runtime, while noting in Table~\ref{tab:all_systems_dist} that its (A0, B0, C0) tuple is closer to \archlp augmentation than to Mooncake-shaped designs; readers who prefer a narrower hybrid-tier definition can read ShadowKV as a \archlp boundary system without changing the rest of the analysis. FlexGen~\cite{sheng2023flexgen} is treated as a local-control tiering precursor in Table~\ref{tab:all_systems_local} rather than as a hybrid-tier system, because its placement search across GPU, CPU, and disk was designed for offline batched generation rather than for the multi-mechanism integration that defines the archetype. Trinity~\cite{liu2025trinity} is related when vector-search disaggregation is coupled to prefill/decode placement, but its primary contribution is not a general multi-tier KV-cache manager; we therefore mark it as a boundary case (see~\S\ref{sec:boundary}). Hybrid-tier systems expose control-plane interactions---coordinator contention, cross-tier eviction ordering, multi-lifetime management, and interactions between handoff and reuse---that single-response systems can avoid.

The defining characteristic of hybrid-tier systems is integration: they occupy multiple values on one or more axes simultaneously and arbitrate among them with explicit policy. Recording that range is more informative than forcing a single-tuple classification, and it suggests that as the field matures, the taxonomy may need an explicitly compositional extension (see~\S\ref{sec:threats}).

\subsection{A Brief Timeline of System Evolution}
\label{sec:timeline}

\begin{figure}[t]
 \centering
 \includegraphics[width=\columnwidth]{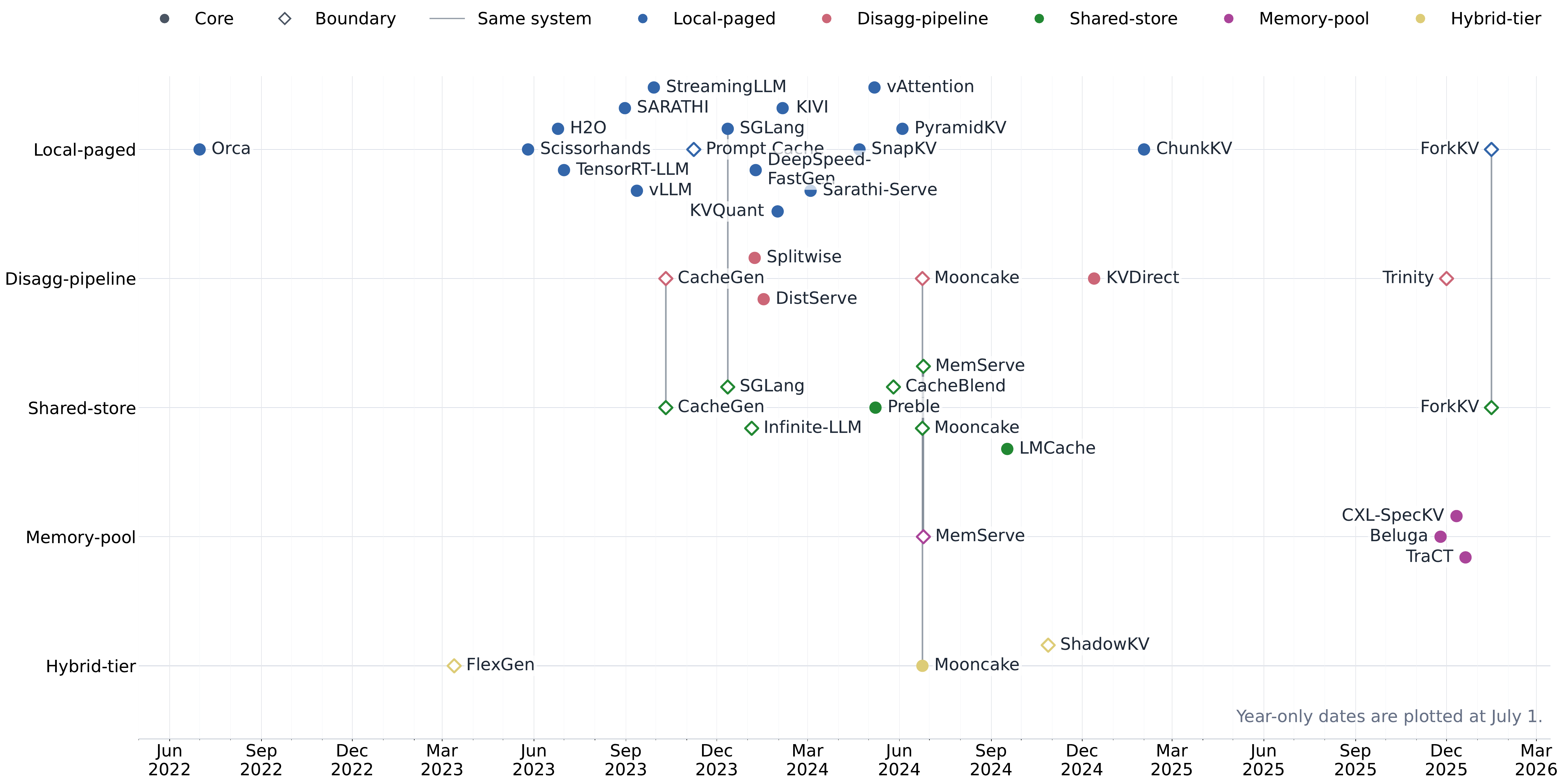}
 \caption{Timeline of LLM serving and KV-cache system designs. Systems are positioned by release date and grouped by the five archetypes used in this survey: local-paged, disaggregated-pipeline, shared-store, memory-pool, and hybrid-tier. Filled circles indicate primary archetype assignments; hollow diamonds indicate boundary or secondary assignments (per~\S\ref{sec:fivearch} and~\S\ref{sec:boundary}). FlexGen and ShadowKV are drawn as boundary diamonds in Hybrid-tier to match their treatment in the text as precursor and edge-of-definition cases respectively; Mooncake is the sole filled hybrid-tier anchor. Boundary systems Prompt Cache, Infinite-LLM, MemServe, and ForkKV are plotted on their respective archetype rows; systems classified under more than one archetype are linked by thin vertical connectors.}
 \label{fig:timeline}
\end{figure}

The chronology helps explain why the five archetypes emerged in parallel rather than as a simple replacement sequence. Figure~\ref{fig:timeline} therefore groups systems by their archetype role, not by the six background responses: KV reduction appears inside \archlp because it keeps ownership local, and offload or tiering appears in \archht when it is composed with disaggregation, reuse, or multi-tier placement. FlexGen is the one precursor we plot in \archht rather than \archlp: although its ownership is local (A0, C0), its defining contribution is GPU/CPU/disk tiered placement, which anticipates the hybrid-tier cost model more directly than the local allocator and scheduler work; Table~\ref{tab:all_systems_local} still records its primary role as a local-control offload precursor.

Local serving mechanisms---iteration-level scheduling, PagedAttention-style block management, chunked prefill, KV eviction, and quantization---were established early and remain active because they improve the local allocator, scheduler, or representation. Disaggregated-pipeline and shared-store systems became prominent in 2024 as prefill/decode interference and repeated-context workloads made one-worker KV ownership less attractive. Memory-pool systems appear later, consistent with the slower availability of CXL-class substrates. Hybrid-tier systems compose multiple archetypes rather than replacing earlier ones and therefore expose interactions among handoff timing, reuse lifetime, tiered placement, and ownership. The publication cycle still creates a comparison gap: these systems are evaluated on different models, context distributions, hardware generations, batching policies, and SLOs, so chronology alone cannot rank them.

\subsection{Within-Archetype Variance}
\label{sec:variance}

Table~\ref{tab:variance} reports how many distinct values each archetype takes on each axis. Few values mean the members truly converge; many values mean the archetype spans diverse design points.

\begin{table}[h]
\small
\centering
\caption{Within-archetype variance per axis. Counts are approximate distinct values because boundary systems may occupy multiple values.}
\label{tab:variance}
\begin{tabular}{@{}lcccc@{}}
\toprule
\textbf{Archetype} & \textbf{A} & \textbf{B} & \textbf{C} & \textbf{D values} \\
\midrule
Local-paged & 1 & 2--3 & 1 & 1--2 \\
Disagg-pipeline & 1 & 1 & 2 & 2 \\
Shared-store & 1 & 2 & 2 & 3--4 \\
Memory-pool & 1 & 2--3 & 1 & 2--3 \\
Hybrid-tier & 3 & 3 & 2 & 3--4 \\
\bottomrule
\end{tabular}
\end{table}

The strongest convergence appears on locality and ownership. Local-paged has a large population and its low A/C variance reflects genuine convergence around per-worker control; memory-pool has similarly stable A/C values but its small population may be a sample-size artifact. Variance is highest in hybrid-tier, which is consistent with the archetype's definition. Shared-store is the telling case: its ownership splits between coordinator designs (C1) and distributed lookup-or-transfer protocols (C2), a choice driven by engineering preference rather than by the substrate.

\subsection{Boundary Systems}
\label{sec:boundary}

Several systems sit between archetypes. \textbf{SGLang} is local-paged in its default radix-tree mode and shared-store-adjacent with HiCache enabled, because KV is then stored outside the immediate GPU-resident block table. \textbf{CacheGen} compresses KV for network transfer rather than defining a persistent shared namespace, so it is closest to disaggregated-pipeline at request scope (B0) and shared-store-adjacent when the compressed representation supports remote reuse. \textbf{KVDirect} preserves the disaggregated-pipeline locality and lifetime but uses peer-to-peer ownership (C2) instead of a coordinator (C1), which is architecturally significant but does not warrant a separate archetype. \textbf{Mooncake}'s store layer resembles \archss, but the complete system composes disaggregation, shared storage, and tiered placement under a larger control plane and is classified as hybrid-tier; the store layer is used as shared-store evidence only when discussing that component.

\textbf{Trinity}~\cite{liu2025trinity} shares the (A1, B0, C1, D-local+D1) signature of DistServe and Splitwise; its boundary label records the contribution type---coupling vector-search retrieval to the PD pipeline---rather than an axis-tuple difference. \textbf{Infinite-LLM}~\cite{lin2024infinitellm} is shared-store-adjacent with the (A2, B0, C2) signature derived in~\S\ref{sec:worked_example}; the request-scoped lifetime distinguishes it from cross-session shared stores. \textbf{Prompt Cache}~\cite{gim2024promptcache} is local-paged-adjacent with (A0, B2+B3, C0): its reuse unit is a structured prompt schema rather than a token prefix, which stresses the B2 lifetime definition without changing locality or ownership. \textbf{ForkKV}~\cite{wang2026forkkv} supplies forking semantics for KV reuse across reasoning branches, with (A0/A2, B0/B2+B3, C0/C2) values that depend on configuration; it is boundary to both \archlp and \archss.

\subsection{Threats to Validity}
\label{sec:threats}

The analysis is empirical rather than exhaustive. The five envelopes describe the classified systems; they are not a proof that no sixth can exist. Because substrate choices constrain locality and ownership, some of the apparent convergence reflects today's hardware. Hybrid-tier systems are better represented as axis ranges than as single tuples, so the within-archetype variance in Table~\ref{tab:variance} is a lower bound on diversity. The memory-pool envelope rests on a small, recent population, so its low locality and ownership variance may be a sample-size artifact that later CXL systems will widen, and future speculative-decoding-aware or MoE-aware KV managers may motivate a compositional extension of the taxonomy.

\subsection{Cross-Archetype Comparison: What the Archetypes Reveal}
\label{sec:crossarchetype}

Among the distributed archetypes, hardware shapes locality and workload shapes lifetime, leaving ownership as the axis on which otherwise-matched systems most often disagree. DistServe and KVDirect share locality, lifetime, and substrate (A1, B0, D-local+D1) but differ on ownership: DistServe uses a centralized goodput-optimized scheduler (C1), while KVDirect embeds a peer-to-peer transfer protocol (C2). Mooncake and LMCache share locality and reuse scope (A2, B2+B3) but differ on ownership: Mooncake routes through a centralized Conductor (C1), while LMCache uses content-derived chunk keys over centralized or distributed backends (C1/C2). The choice among coordinator, distributed-protocol, and shared-memory ownership carries scalability and fault-tolerance consequences that end-to-end latency hides. Ownership is not universally dominant; it is simply where the distributed systems in this population diverge most.

Substrate choices, by contrast, are converging in practice: most surveyed systems use GPU HBM for active KV, RDMA-capable networking for cross-node movement, host DRAM for offload, and increasingly CXL for pooled designs. For emerging memory-pool systems, the classification indicates architectural role rather than validated comparative performance, and future systems may differentiate less on the substrate itself and more on the ownership and lifetime policies built over it.

The empty B4 region is an important absence: no surveyed system provides durable KV state with an explicit recovery model for node or store failure. This is distinct from semantic response caching or retrieval caching~\cite{regmi2024gptcache}; durable KV would require preserving, validating, and recovering model-specific KV tensors or blocks. The absence becomes costlier as KV state grows to tens or hundreds of gigabytes per active workload, because a lost store mid-decode forces re-prefill, slow-path fallback, or request abort.

A less visible second gap sits at $(A3, B2+B3, C2/C3, D3)$: a CXL-backed prefix cache with cross-session reuse, opportunistic retention, and explicit ownership semantics. The constituent pieces already exist---CXL shared KV paths in memory-pool systems, distributed prefix directories in shared-store systems---but the combination has not been demonstrated as a documented serving architecture. \S\ref{sec:agenda} turns both this absence and the B4 gap into concrete open questions.

\section{Analysis of LLM Serving Systems}
\label{sec:analysis}

Section~\ref{sec:archetypes} named five archetypes; this section provides the underlying evidence. We classify the surveyed systems on the four axes (Tables~\ref{tab:all_systems_local}~and~\ref{tab:all_systems_dist}), compare six representative anchors on recurring mechanisms (Table~\ref{tab:mechanisms}), and audit which evaluations report which KV-specific measurements (Table~\ref{tab:gaps}). The audit yields seven measurement gaps that motivate the agenda in~\S\ref{sec:agenda}. The tables mix peer-reviewed systems, preprints, and production frameworks; the classification records architectural semantics, not a normalized performance ranking.

\subsection{System Classification}
\label{sec:systemtable}

\begin{table*}[t]
\small
\centering
\caption{System classification --- local and framework-centered systems. A=locality, B=lifetime, C=ownership, and Substrate=D-level memory or transfer path; anchor systems are in \textbf{bold}. The Tier column records evidence type: P=peer-reviewed venue, A=arXiv preprint, V=vendor or industry technical report, W=workshop venue, F=open-source framework.}
\label{tab:all_systems_local}
\resizebox{\textwidth}{!}{%
\begin{tabular}{@{}llccccp{2cm}p{4.5cm}@{}}
\toprule
\textbf{System} & \textbf{Role} & \textbf{A} & \textbf{B} & \textbf{C} & \textbf{Tier} & \textbf{Substrate} & \textbf{Primary KV Contribution} \\
\midrule
\textbf{vLLM} & Local-paged & A0 & B0/B2+B3 & C0 & P & D-local & PagedAttention block-level allocation; continuous batching and block reuse \\
SGLang (local) & Local-paged & A0 & B2+B3 & C0 & P & D-local & RadixAttention prefix sharing; instance-local opportunistic reuse \\
\textbf{Sarathi-Serve} & Local-paged & A0 & B0 & C0 & A & D-local & Chunked prefill; stall-free scheduling \\
Orca & Local-paged & A0 & B0 & C0 & P & D-local & Iteration-level scheduling; preemptive batching \\
SARATHI & Local-paged & A0 & B0 & C0 & A & D-local & Chunked prefill; piggybacking decodes with prefills \\
H2O & Local-paged & A0 & B0 & C0 & P & D-local & Heavy-hitter KV eviction; attention-score-based retention within a request \\
Scissorhands & Local-paged & A0 & B0 & C0 & P & D-local & KV persistence hypothesis; fixed-budget retention within a request \\
StreamingLLM & Local-paged & A0 & B0 & C0 & P & D-local & Attention sinks; streaming-session retention with limited window \\
SnapKV & Local-paged & A0 & B0 & C0 & P & D-local & Prompt-aware KV compression before generation \\
PyramidKV & Local-paged & A0 & B0 & C0 & P & D-local & Layer-wise asymmetric KV budget allocation \\
KIVI & Local-paged & A0 & B0 & C0 & P & D-local & 2-bit asymmetric KV quantization \\
KVQuant & Local-paged & A0 & B0 & C0 & P & D-local & Non-uniform KV quantization; 1M context on single GPU \\
ChunkKV & Local-paged & A0 & B0 & C0 & A & D-local & Semantic-chunk-based KV compression \\
DeepSpeed-FastGen & Local-paged & A0 & B0 & C0 & V & D-local & SplitFuse; fused prefill-decode chunk scheduling \\
Prompt Cache & Local-paged boundary & A0 & B2+B3 & C0 & P & D-local & Schema-aware prompt KV caching at semantic-template granularity \\
FlexGen~\cite{sheng2023flexgen} & Local-offload precursor & A0 & B0 & C0 & P & D-local,D4,D5 & Single-GPU offload of inference state across GPU, CPU, and disk \\
\midrule
TensorRT-LLM & Framework & A0 & B0/B2+B3 & C0 & F & D-local,D0 & Production paged KV cache, in-flight batching, and optional KV reuse/offload support \\
vAttention & Framework & A0 & B0 & C0 & A & D-local & Virtual memory for KV; CUDA virtual address management \\
\midrule
\multicolumn{8}{l}{\footnotesize Role records the dominant role in this survey. D-local denotes ordinary GPU HBM/local device memory.}
\end{tabular}
}
\end{table*}

\begin{table*}[t]
\centering
\caption{System classification---distributed, pooled, and hybrid systems.
A=locality, B=lifetime, C=ownership, and D=memory or transfer substrate.
Anchor systems are shown in \textbf{bold}.
The Tier column records evidence type: P=peer-reviewed, A=arXiv, V=vendor, W=workshop, F=framework.}
\label{tab:all_systems_dist}

\scriptsize
\setlength{\tabcolsep}{2.2pt}
\renewcommand{\arraystretch}{1.15}

\begin{tabularx}{\textwidth}{@{}L{1.85cm}L{1.65cm}cL{0.95cm}L{0.85cm}cL{1.7cm}Y@{}}
\toprule
\textbf{System} &
\textbf{Role} &
\textbf{A} &
\textbf{B} &
\textbf{C} &
\textbf{Tier} &
\textbf{Substrate} &
\textbf{Primary KV Contribution} \\
\midrule

\textbf{DistServe} &
Disagg. pipeline &
A1 &
B0 &
C1 &
P &
D-local, D1 &
PD disaggregation with goodput-optimized scheduling \\

\textbf{Splitwise} &
Disagg. pipeline &
A1 &
B0 &
C1 &
P &
D-local, D1 &
Phase-aware allocation with separate prefill/decode clusters \\

KVDirect &
Disagg. pipeline &
A1 &
B0 &
C2 &
A &
D-local, D1 &
Distributed KV transfer protocol and peer-to-peer handoff \\

\midrule

\textbf{LMCache} &
Shared store &
A2 &
B2+B3 &
C1/C2 &
A &
D-local, D4, D5 &
Distributed KV object store for cross-instance prefix reuse \\

\textbf{Preble} &
Shared store &
A2 &
B2+B3 &
C1 &
A &
D-local, D1 &
Global scheduler with prefix-aware KV placement \\

SGLang HiCache &
Shared-store adj. &
A2 &
B2+B3 &
C1/C2 &
F &
D4, D5 &
Hierarchical prefix cache beyond one serving instance \\

CacheBlend~\cite{yao2024cacheblend} &
Shared-store adj. &
A2 &
B2+B3 &
C1/C2 &
P &
D-local, D4, D5 &
Cached knowledge fusion for reusable RAG context \\

Infinite-LLM~\cite{lin2024infinitellm} &
Shared-store adj. &
A2 &
B0 &
C2 &
A &
D-local, D1 &
Cluster-level KV pool for long-context distributed attention \\

CacheGen &
Boundary &
A1/A2 &
B0/B2 &
C0/C1 &
P &
D1, D4, D5 &
KV compression and streaming for transfer or context loading \\

Mooncake store &
Shared-store comp. &
A2 &
B2+B3 &
C1 &
A &
D-local, D1, D4, D5 &
KVCache-centric disaggregation with a tiered KV store \\

\midrule

\textbf{TraCT} &
Memory pool &
A3 &
B0+B2+B3 &
C3 &
A &
D-local, D3 &
Rack-scale CXL shared-memory KV with load/store access \\

Beluga~\cite{yang2025beluga} &
Memory pool &
A3 &
B0/B2+B3 &
C3 &
A &
D-local, D3 &
CXL memory pooling with a KV-aware architecture \\

CXL-SpecKV~\cite{liu2025cxlspeckv} &
Memory pool &
A3 &
B0+B3 &
C3 &
W &
D3, D6 &
CXL-disaggregated speculative KV cache with FPGA-side support \\

\midrule

\textbf{Mooncake} &
Hybrid tier &
A1+A2 &
B0+B2+B3 &
C1 &
A &
D-local, D1, D4, D5 &
PD disaggregation, distributed KV storage, and tiering \\

\textbf{ShadowKV} &
Hybrid tier (boundary) &
A0 &
B0 &
C0 &
A &
D-local, D4 &
Local-control KV offload with sparse retrieval and reconstruction \\

Trinity &
Boundary &
A1 &
B0 &
C1 &
A &
D-local, D1 &
Vector-search disaggregation coupled to PD placement \\

ForkKV &
Boundary &
A0/A2 &
B0/B2+B3 &
C0/C2 &
A &
D-local, D1 &
Forking semantics for branching/agentic KV reuse \\

\midrule
\multicolumn{8}{@{}p{\dimexpr\textwidth-2\tabcolsep\relax}@{}}{\scriptsize
\textit{Notes:} Role records the dominant role in this survey.
D1=RDMA/GPUDirect RDMA; D3=CXL-like shared memory; D4=host DRAM;
D5=SSD/storage; D6=SmartNIC/DPU/FPGA.} \\
\bottomrule
\end{tabularx}
\end{table*}

Tables~\ref{tab:all_systems_local}~and~\ref{tab:all_systems_dist} together classify 35 entries spanning 33 distinct systems and framework modes (18 entries in Table 8, 17 in Table 9); SGLang and Mooncake each appear in two modes, which is why entries exceed systems. Local-paged is the largest archetype because compression, eviction, virtual memory, and scheduling work can be evaluated on a single serving instance against a mature PagedAttention-style baseline. Disagg-pipeline and shared-store differ primarily in lifetime: the former transfers request-scoped KV; the latter retains it across requests under opportunistic eviction. Hybrid-tier and boundary entries expose composition---Mooncake combines disaggregation, shared storage, and tiering; ShadowKV combines tiering with sparse retrieval; Trinity~\cite{liu2025trinity} couples vector-search placement to the PD pipeline without making reusable KV state its central managed object. The evidence-tier column reminds the reader that several headline systems are arXiv or framework sources; their performance numbers should be read as author-reported until reproduced.

\subsection{Mechanism-Level Comparison}
\label{sec:mechanisms}

\begin{table*}[t]
\centering
\caption{Mechanism comparison across representative anchor systems.
Entries summarize the dominant mechanism reported or implied by each anchor;
``---'' means that the mechanism is absent or not specified as a KV-management feature.}
\label{tab:mechanisms}

\scriptsize
\setlength{\tabcolsep}{2.6pt}
\renewcommand{\arraystretch}{1.15}

\begin{tabularx}{\textwidth}{@{}L{2.15cm}YYYYYY@{}}
\toprule
\textbf{Mechanism} &
\textbf{vLLM} &
\textbf{DistServe} &
\textbf{LMCache} &
\textbf{Mooncake} &
\textbf{TraCT} &
\textbf{ShadowKV} \\
\midrule

Block allocation &
Paged\-Attention &
Paged\-Attention-like &
Paged\-Attention-like &
Paged\-Attention-like &
CXL-aware layout &
Offload-aware layout \\

Naming / prefix ID &
Block table &
Request/job ID &
Content-derived keys &
Conductor-managed &
Pool metadata &
Request/context ID \\

Transfer/access &
Local HBM &
RDMA push/pull &
Remote fetch &
RDMA push/pull &
CXL load/store &
CPU offload and reconstruction \\

Readiness signal &
Local scheduler &
Handoff completion &
Object/block readiness &
Block/tier readiness &
Allocator metadata &
Prefetch readiness \\

Eviction &
Local block policy &
Request completion &
Store policy &
Tier-aware policy &
Pool allocator &
Sparse retrieval policy \\

Multi-tenant isolation &
--- &
--- &
Namespace &
--- &
--- &
--- \\

Fault tolerance &
--- &
--- &
--- &
--- &
--- &
--- \\

\bottomrule
\end{tabularx}
\end{table*}

Table~\ref{tab:mechanisms} compares six representative anchor systems across the mechanisms introduced in~\S\ref{sec:projections}. Two mechanisms recur across archetypes: many anchors use PagedAttention-style block allocation or an equivalent block layout, and representative systems that cross a network commonly use RDMA-capable push, pull, or fetch paths. CacheGen can use TCP as a non-anchor boundary mechanism, but the high-performance distributed anchors are designed around lower-latency transfer or shared-memory substrates.

The remaining mechanisms vary, and the variance tracks the taxonomy. Naming tracks ownership: vLLM uses request-local tables, Mooncake delegates naming and placement to its centralized \emph{Conductor}, and LMCache uses content-derived keys. Readiness signaling tracks lifetime: B0 systems can use request-scoped completion because KV does not outlive the request, while B2+B3 systems need object-, block-, or tier-granularity readiness signals to support overlap and opportunistic retention. Eviction policies are usually described functionally but rarely measured as a control-plane cost; no representative anchor characterizes reuse-distance distributions, which is the strongest single piece of evidence for MG4 in~\S\ref{sec:agenda}. Multi-tenant isolation and fault tolerance are absent or only partially addressed across the representative anchors, which is why they reappear as DG2 and DG1.

A pattern not visible in the archetype table is that completion signaling changes as soon as KV crosses a locality or lifetime boundary: local systems use per-request synchronization, disagg-pipeline systems need handoff completion, shared-store and hybrid-tier systems need block- or tier-granularity readiness signals, and memory-pool systems inherit coherence or allocator-level synchronization. Completion signaling is therefore a convergent control-plane primitive even when data-plane substrates differ, which is why the most reusable primitives across archetypes are naming, readiness, eviction, and completion rather than the data path itself.

\subsection{Cross-Cutting Observations from System Classification}
\label{sec:crosspatterns}

Compression and eviction systems cluster on the local-control side, while distributed systems cluster around placement, movement, and ownership: KIVI, KVQuant, ChunkKV, H2O, Scissorhands, StreamingLLM, SnapKV, PyramidKV, GEAR~\cite{kang2024gear}, DynamicMemCompress~\cite{nawrot2024dynamic}, and ZipCache~\cite{he2024zipcache} optimize representation; distributed systems decide which device, worker, or tier holds each block. Systems such as CacheGen and CXL-SpecKV combine the two, but one side is usually the primary contribution. Locality and lifetime are correlated but not identical: A0 systems mostly use request-scoped KV, while A2 and A3 designs more often appear with cross-request reuse (B2) or opportunistic retention (B3) because moving KV across a boundary must buy reuse, capacity, or transfer overlap. Substrate diversity increases with archetype complexity---local HBM in \archlp, RDMA in \archdp, host memory or storage in \archss, CXL in \archmp, and several substrates composed in \archht---which is why ownership becomes more important in later archetypes: each additional substrate requires explicit scheduling, coordination, or consistency machinery. Publication maturity does not map cleanly onto this progression, so all five archetypes remain active rather than forming a replacement sequence.

\subsection{A Testable Prediction from the Taxonomy}
\label{sec:prediction}

A taxonomy is most useful when it yields a prediction whose failure would force a revision. Ours suggests one. Among the distributed archetypes (\archdp, \archss, and the distributed components of \archht), at equivalent throughput and matched workload, systems with centralized ownership (C1) should show lower tail-latency variance than systems with distributed ownership (C2), because C1 removes distributed-coordination stalls from the critical path while paying a fixed scheduling cost. Concretely: a 99th-percentile TTFT decomposition for a C1 system should be dominated by queueing and prefill compute, while a comparable decomposition for a C2 system should additionally show contributions from directory lookup, distributed-lock contention, and consensus-style synchronization. DistServe versus KVDirect is the natural test pair because they hold $A$, $B$, and $D$ fixed and differ on $C$ alone; Mooncake versus LMCache provides a second test pair in the shared-store envelope. We cannot fully evaluate this prediction with currently reported data---the tail decomposition is exactly MG6, the gap surfaced by the audit---which itself reinforces the case for the measurement program. If a future study found indistinguishable tail decompositions for C1 and C2 systems on matched workloads, the taxonomy would have to demote ownership or add a confounding axis. The prediction is offered as a falsifiable consequence, not an asserted result.

\subsection{Measurement-Gap Audit}
\label{sec:gaps}

Most systems report at least one end-to-end metric---throughput, TTFT, TPOT, goodput, or tail latency---but the exact metric set differs by archetype and workload. In Table~\ref{tab:gaps}, a metric is counted as reported only when the paper gives it in a form that can support cross-system comparison or root-cause attribution, not merely when the term appears in an evaluation section. Table~\ref{tab:gaps} shows that seven KV-specific measurements are rarely reported in such a form. We audit them in turn.

\begin{table*}[t]
\small
\centering
\caption{Measurement-gap audit. A metric is counted as ``reported'' (\cmark) only when the paper presents it in a form that supports cross-system comparison or root-cause attribution; ``partial'' means some systems report it under restrictive conditions (microbenchmark only, single substrate only, or single workload); \xmark\ means no surveyed system in the archetype reports it; N/A means the metric does not apply to the archetype's locality and lifetime pattern.}
\label{tab:gaps}
\resizebox{\textwidth}{!}{%
\begin{tabular}{@{}lccccc@{}}
\toprule
\textbf{Metric} & \textbf{Local-paged} & \textbf{Disagg-pipeline} & \textbf{Shared-store} & \textbf{Memory-pool} & \textbf{Hybrid-tier} \\
\midrule
TTFT / TPOT & partial & \cmark & partial & partial & \cmark \\
Throughput / Goodput & \cmark & \cmark & partial & partial & \cmark \\
MG1: Remote KV access patterns & N/A & \xmark & \xmark & \xmark & \xmark \\
Transfer/access latency & N/A & partial & partial & partial & partial \\
Cache hit rate & N/A & N/A & \cmark & \xmark & \cmark \\
MG2: Metadata path costs & \xmark & \xmark & \xmark & \xmark & \xmark \\
MG3: Completion overhead & \xmark & \xmark & \xmark & \xmark & \xmark \\
MG4: Lifetime / reuse distance & \xmark & \xmark & \xmark & \xmark & \xmark \\
MG5: Prefetchability slack & \xmark & \xmark & \xmark & \xmark & \xmark \\
MG6: P99 tail attribution & \xmark & partial & \xmark & partial & partial \\
MG7: Public trace availability & \xmark & partial & partial & \xmark & partial \\
\bottomrule
\end{tabular}
}
\end{table*}

The missing data-path measurements are MG1 and MG5. Systems that move KV rarely report the event distribution underneath their end-to-end numbers: bytes moved per token, remote blocks touched per token, operation sizes, or critical-path stalls (MG1). Transfer or access latency sometimes appears for a single microbenchmark, but not in a form that can explain P99 TTFT or TPOT. Prefetchability slack (MG5) is missing entirely, even though layer order and token order make many KV accesses predictable.

The missing metadata-path measurements are MG2, MG3, and MG4. Block lookup, prefix matching, ownership tracking, reference counting, eviction decisions, location directories, and completion notifications consume CPU cycles and memory bandwidth, but no system measures their cost (MG2, MG3). Reuse-distance and block-lifetime distributions are also absent (MG4), which leaves eviction policies tuned by intuition rather than workload evidence.

\textbf{Why the metadata gap matters: a back-of-the-envelope.} Consider a shared-store with a radix-tree prefix index holding $N$ active prefixes. Each lookup performs $O(\log N)$ pointer chases, each chase missing the last-level cache at roughly 100~ns. For $N=10^4$ prefixes a lookup costs $\approx 1.3~\mu$s; under a request rate of $10^5$~req/s the lookup path alone consumes about 130~ms of one core every second, before reference counting, copy-on-write tracking, eviction decisions, or replication metadata. Layer-granularity completion signaling in pipelined disaggregation adds a synchronization term that scales with the layer count $L$, while request-granularity completion has a single synchronization point but pins the consumer for the full transfer; the variance contributed by the two strategies is empirically different but unreported. These numbers show the metadata-path gap is substantive, not cosmetic: without it, a shared-store that scales to tens of thousands of prefixes looks identical to one that does not.

MG6 is a cross-cutting attribution gap: decomposing P99 TTFT or TPOT into queueing, compute, transfer, metadata, and synchronization components requires both data-path and metadata-path visibility, which is why it is reported only partially and only for some distributed archetypes. MG7, the public-trace gap, has narrowed but not closed: Splitwise released the Azure LLM Inference traces (with request arrivals, prompt and output lengths, and timing fields), and Mooncake released a production KV-aware trace; both are now widely reused. Neither, however, exposes per-block KV access events, prefix-sharing identities, session boundaries, or transfer-event timings at the granularity needed to attribute P99 tail latency to a specific KV mechanism. Most other surveyed systems still evaluate on synthetic workloads or private traces with unknown prefix-sharing structure. A useful next-generation public trace would include request arrivals, prompt and output lengths, prefix-sharing structure, session structure, KV block accesses, and transfer events. Current evaluations are sufficient to show that a system improves over its own baseline but insufficient to explain why or whether the same mechanism would win under a different workload, SLO, or hardware substrate.

\subsection{Implications for Cross-Archetype Comparison}
\label{sec:crosscomp}

Cross-archetype comparison is currently method-bound: a disaggregated-pipeline paper reporting ``$2\times$ goodput'' and a shared-store paper reporting ``$5\times$ throughput'' rarely share a workload, an SLO, or a hardware family, so comparisons between them are necessarily qualitative without a common trace and KV-bytes-moved counters. Individual evaluations are consistent with each system's stated goal; the shortfall is community-level. Speculative decoding, MoE, multi-tenant isolation, and fault tolerance remain weakly covered by the anchor systems: vLLM partially engages speculative decoding and LMCache exposes namespace mechanisms for isolation, but no anchor provides a full KV-management treatment of these cases, which motivates DG2, DG4, and DG5 in~\S\ref{sec:agenda}. For practitioners, \archlp is the natural baseline when KV does not cross workers; \archdp targets phase interference; \archss requires enough prefix or context reuse to justify lookup and transfer overhead; \archmp is most relevant when CXL-like fabrics are available; \archht matches production stacks but is hardest to evaluate because several control planes interact.
\section{Research Agenda}
\label{sec:agenda}

The audit in~\S\ref{sec:gaps} surfaced what current evaluations omit. The agenda pairs seven design gaps (DG1--DG7) with seven measurement gaps (MG1--MG7); each design question depends on a specific subset of missing counters, traces, or decompositions, so measurement is a precondition for architecture rather than an afterthought. The gaps are KV-specific---block lifetime, prefix reuse distance, transferred-KV completion, remote KV access distributions, and metadata operations such as block lookup, reference counting, and location directories---rather than generic distributed-systems concerns.

\subsection{Design Gaps}

\textbf{DG1---Fault tolerance of distributed KV state.} When a distributed-KV store fails mid-decode, should the system re-prefill from the original prompt, recover from another tier or replica, or fail the request? This is distinct from semantic response caching: the object to recover is model-specific KV tensor state whose validity depends on model version, tokenization, position encoding, adapters, and attention layout. The options have very different latency and cost profiles, but no surveyed system publishes a KV recovery model. MG3 and MG4 are prerequisites: completion semantics identify which blocks became visible to consumers, while lifetime distributions identify how long recoverable state must survive.

\textbf{DG2---Multi-tenant isolation.} A shared prefix cache can leak information because a hit is observably faster than a miss and may reveal that a prior tenant used the same prefix. The open question is what a timing-channel-aware prefix cache should guarantee, and what latency benefit remains after defenses such as namespace isolation, admission rules, padding, or noise. MG6 and MG7 are prerequisites because the defense cost depends on tail-latency attribution and multi-tenant reuse structure.

\textbf{DG3---Eviction across replicated tiers.} A KV block can be live in HBM, host DRAM, and a remote store simultaneously. Existing tier-aware systems often approximate this with per-tier heuristics. A principled policy would price each tier's fetch latency against the probability of near-future reuse, which requires per-block reuse-distance and lifetime traces. MG4 and MG5 are therefore prerequisites.

\textbf{DG4---Speculative decoding $\times$ KV cache.} Speculative decoding creates tentative KV columns that may be accepted or rolled back. Existing distributed-KV systems mostly assume linear append, so they do not specify whether branch KV should be transferred, shared, or discarded before verification. RDMA-based transfer and CXL-style shared memory have different rollback costs; MG1 and MG3 are needed to model the resulting trade-off.

\textbf{DG5---MoE $\times$ KV cache.} In standard Transformer MoE models, attention KV follows the attention-layer layout rather than expert identity, but expert all-to-all traffic still competes with KV transfer and prefetch traffic. The missing design question is whether expert routing and KV placement should be co-scheduled or merely share a congestion-aware runtime. MG1 and MG2 are prerequisites because the answer depends on joint data-path and metadata-path traffic.

\textbf{DG6---Ownership semantics for shared prefix caches.} Shared prefix caches rely on reference counting, copy-on-write, and location directories, but rarely state a consistency contract. If one request holds prefix $P$ while another extends or evicts it, the system should define what each consumer can observe. MG2 is prerequisite because the cost of stronger semantics depends on which metadata operations are frequent, synchronized, or on the critical path.

\textbf{DG7---Per-session KV management.} The B1 lifetime level (\S\ref{sec:taxonomy}) is currently unoccupied by published systems: deployed chat services retain per-session KV state across conversational turns, but no surveyed system formalizes a per-session retention contract distinct from streaming-window mechanisms or instance-local opportunistic reuse. The open question is what a published per-session KV architecture would specify---retention budgets per session, eviction priorities across active sessions, migration semantics when a session moves between workers, and isolation between sessions of different tenants. MG4 (reuse-distance distributions) and MG7 (session-aware public traces) are prerequisites because a session-aware policy needs evidence about inter-turn arrival gaps and reuse patterns.

\subsection{Measurement Gaps}

\textbf{MG1---Remote KV access-pattern characterization.} Systems report TTFT and TPOT but not the remote-memory event distribution: operation sizes, remote blocks per token, remote bytes per token, and critical-path remote stalls. These counters are what shared-store and hybrid-tier systems actually wait on.

\textbf{MG2---Metadata path costs.} Systems should report how much time and CPU/memory bandwidth are spent on block lookup, prefix lookup, ownership tracking, refcounting, eviction decisions, location directories, and completion notifications. Without this breakdown, DG5 and DG6 cannot distinguish data-path bottlenecks from control-plane bottlenecks.

\textbf{MG3---Completion and synchronization overhead.} Distributed KV systems need to know when a block, layer, or request is safe to consume. The missing measurement is the latency cost and correctness trade-off of each completion granularity, especially under asynchronous transfer.

\textbf{MG4---Reuse-distance and lifetime distributions.} This measurement records how long KV blocks live and how far apart their reuses are. Aggregate hit rate does not reveal whether a block should be replicated, offloaded, or evicted.

\textbf{MG5---Prefetchability slack.} Transformer inference exposes future layer and token access order, but systems do not report how much time exists between a predictable need and the deadline for that block. This quantity determines whether tiered KV policies can rely on prefetch.

\textbf{MG6---Tail-latency attribution.} This measurement decomposes P99 TTFT or TPOT into queuing, compute, transfer, metadata, and synchronization components. Without attribution, optimization efforts may target the wrong bottleneck; DG2 also needs it because timing channels are visible at the tail, not just in the mean.

\textbf{MG7---Public traces.} A KV-aware trace should include request arrivals, prompt/output lengths, prefix-sharing structure, session structure, KV block accesses, and transfer events. Mooncake's production trace is a notable exception; most systems still rely on synthetic or private workloads.

\subsection{Pairings}

Table~\ref{tab:pairings} pairs each design gap with the specific measurements needed to address it. The pairing is the argument that follow-up measurement work is well-motivated: without MG4 (lifetime distributions), fault-tolerance mechanisms (DG1) cannot be designed against a known state-retention target.

\begin{table}[h]
\small
\centering
\caption{Pairing design gaps with required measurements. DG denotes an open design gap; MG denotes a missing measurement needed to evaluate or design that mechanism.}
\label{tab:pairings}
\begin{tabular}{@{}p{3.8cm}p{9.7cm}@{}}
\toprule
\textbf{Design gap} & \textbf{Required measurement} \\
\midrule
DG1: Fault tolerance & MG3 (completion semantics) + MG4 (lifetime distributions) \\
DG2: Multi-tenant isolation & MG6 (tail-latency attribution) + MG7 (multi-tenant traces) \\
DG3: Replicated-tier eviction & MG4 (reuse distance) + MG5 (prefetchability) \\
DG4: Speculative $\times$ KV & MG1 (access patterns) + MG3 (completion) \\
DG5: MoE $\times$ KV & MG1 (access patterns under expert routing) + MG2 (metadata path under EP) \\
DG6: Ownership semantics & MG2 (metadata path) \\
DG7: Per-session KV management & MG4 (reuse distance) + MG7 (session-aware traces) \\
\bottomrule
\end{tabular}
\end{table}

The precondition is short. Closing MG3 (completion semantics) and MG4 (reuse-distance distributions) unblocks fault tolerance (DG1), and MG4 is additionally an input to DG3 and DG7. A single public production trace annotated with reuse distance and tail latency supplies inputs that several questions share, because MG4, MG6, and MG7 can be derived from the same trace. We therefore prioritize completion semantics, reuse-distance distributions, and public KV-aware traces, in that order.
\section{Related Work}
\label{sec:related}

No prior survey covers the four-axis vocabulary or the archetype analysis presented in~\S\ref{sec:taxonomy}--\S\ref{sec:archetypes}, but several existing surveys cover overlapping ground. We position each by the design question its taxonomy answers and the design question it cannot.

\subsection{KV Cache Surveys}

Miao et al.~\cite{miao2026servingsurvey} survey generative LLM serving broadly: batching, scheduling, parallelism, memory management, and deployment stacks. Their taxonomy explains how serving systems improve throughput and latency, but it does not distinguish two KV-specific designs with the same serving goal and different ownership semantics, such as Mooncake's Conductor-managed placement and LMCache's content-addressed chunk lookup. Our axes make that distinction explicit. Their survey covers a wider range of infrastructure than this paper but treats each area as a separate category rather than as a set of interacting design decisions. Our survey is narrower---KV-cache management rather than the whole serving stack---but deeper on the interactions between locality, lifetime, ownership, and substrate.

Li et al.~\cite{li2024survey} organize KV-cache management vertically: token-level selection and dropping, model-level quantization and low-rank decomposition, and system-level hardware-aware design. Their taxonomy captures compression mechanisms in detail, yet DistServe and KVDirect land in the same system-level bucket despite making opposite ownership choices. Our four axes cut across their levels and recover the distinction theirs collapses. Li et al. also provide the most comprehensive coverage of KV compression techniques (12+ papers), which complements our survey: we treat compression as a data-path optimization and focus on the control plane.

Liu et al.~\cite{liu2025kvcachereview} review KV-cache compression for inference efficiency, including selective-token strategies, quantization, and attention compression. Their compression coverage is complementary to Li et al.'s management-oriented view. We classify representative KV compression and eviction systems because they change the capacity and movement cost model, but we do not attempt to re-survey the algorithmic details of every compression method.

\subsection{LLM Inference Efficiency Surveys}

Tay et al.~\cite{tay2022efficient} survey efficient Transformer architectures including sparse and linearized attention. Useful as it is for model-level efficiency, the taxonomy has no vocabulary for whether a KV block is local, remote, shared, or durable—the very thing that defines control-plane systems like LMCache and Mooncake.

Adjacent literatures shape the design space without joining the classified population: general LLM serving systems studying placement, multiplexing, migration, and admission~\cite{li2023alpaserve,wu2023fastserve,sun2024llumnix,duan2024muxserve,aminabadi2022deepspeedinference,song2023powerinfer,shi2025nexus,zhang2025jenga}; long-context, sequence-modeling, and KV-reduction techniques that change the amount or pattern of KV state to manage~\cite{chen2023longlora,ding2024longrope,mohtashami2023landmark,han2023lminfinite,gu2023mamba,dao2023hyena,sun2023retnet,jacobs2023deepspeedulysses,child2019sparse,gu2022efficiently,adnan2024keyformer,lesens2025kqsvd,wang2024adaptivekv,zhang2024adakv,tang2024quest,zhang2024razorattention,ribar2023sparq,liu2024minicache,xu2024less,yang2024pyramidinfer,nawrot2024dynamic,he2024zipcache}; and speculative decoding, MoE, memory disaggregation, retrieval, and vector search, which provide boundary conditions through non-linear lifetimes, shared-fabric contention, remote-memory substrates, or reusable external context~\cite{li2024eagle,fu2024lookahead,xu2024anpd,shazeer2017outrageously,rajbhandari2022deepspeedmoe,hwang2022tutel,gale2022megablocks,lepikhin2020gshard,du2021glam,he2022semoe,he2021fastmoe,ousterhout2010ramcloud,dragojevic2014farm,gu2017infiniswap,shan2018legoos,ruan2020aifm,wang2020semeru,weiner2022tmo,gao2023dracksim,lee2019latent,guu2020realm,karpukhin2020dpr,lewis2020rag,izacard2022atlas,khattab2020colbert,malkov2018hnsw,johnson2019billion}.

\subsection{Memory Disaggregation and CXL Surveys}

Wang et al.~\cite{li2024disaggregation} survey memory disaggregation for data centers, covering RDMA-based and CXL-based approaches across the hardware, architecture, and software-system layers. Their taxonomy answers a substrate question: how memory is exposed across machines. KV caching adds workload structure that generic disaggregation taxonomies do not model: read-heavy decode, predictable layer order, block tables, and prefix identity. Their survey is essential reading for understanding the hardware evolution that enables the hybrid-tier and memory-pool archetypes, but it does not address the workload-specific policies (prefix naming, eviction, completion signaling) that are the focus of our analysis.

Chen et al.~\cite{shan2024cxl} survey CXL-based computing systems from single-machine memory expansion to distributed shared memory. Their CXL focus explains why memory pooling is becoming feasible, but a CXL pool used as raw shared memory and a CXL-backed prefix cache with distributed ownership read the same in their framing. That difference is Axis C. Their survey supplies the hardware background; our ownership axis supplies the software vocabulary CXL builders need to describe what they have built on top of it.

\subsection{What This Survey Adds}

No prior survey takes locality, lifetime, ownership, and substrate as primary axes: Miao et al.\ organize by serving stack, Liu et al.\ by compression technique, Li et al.\ by KV-management level, Wang et al.\ by memory-disaggregation layer, Tay et al.\ by efficient-attention architecture, and Chen et al.\ by CXL substrate. Each orientation is internally consistent, yet none separates Mooncake from LMCache---two shared-store designs that differ in how they name, place, and locate reusable KV. The four-axis tuple surfaces that distinction; the measurement-gap audit and paired agenda then identify which evaluations omit which inputs and which omissions block which next steps.

The scope boundary also differs: speculative decoding and MoE serving appear at the subset that touches KV management; KV-specific compression and eviction appear when they change serving memory behavior; model-weight compression~\cite{frantar2023gptq,lin2023awq,dettmers2023spqr,xiao2023smoothquant,dettmers2022llmint8} and attention-algorithm work without KV pressure are excluded. No prior survey provides a measurement-gap audit of the kind in Table~\ref{tab:gaps}; the closest precedent is the evaluation-methodology discussion in Miao et al., which identifies serving metrics without enumerating what they fail to capture for distributed KV state.
\section{Conclusion}
\label{sec:conclusion}

Recent KV-cache systems are not a scatter of unrelated point solutions. Local virtualization and scheduling, KV reduction, disaggregation, prefix reuse, offload tiering, and memory pooling answer the same four questions: where the KV cache lives, how long it survives, who owns it, and what substrate carries or exposes it. The taxonomy made those questions explicit; the archetype map and the system tables supplied the evidence. The classified systems fall into five envelopes. Convergence is tightest in \archlp; \archss is the most revealing, since ownership there separates coordinator-managed stores from content-addressed designs like LMCache's chunk keys.

Ownership carries the most engineering leverage among the distributed archetypes. Hardware fixes locality, deployment scale fixes substrate, and workload fixes lifetime; what remains for otherwise-similar systems to disagree on is ownership---Mooncake routes through a centralized Conductor, while LMCache uses content-derived chunk keys and can run over centralized or distributed backends, and the two pay different scalability and fault-tolerance prices for how they name, locate, and place reusable KV. End-to-end metrics do not expose this: a goodput improvement reports better performance on one workload and SLO without identifying whether the gain came from scheduling, KV transfer, prefix hits, prefetching, or lower metadata cost. Recent hybrid systems---Mooncake, ShadowKV, Trinity, and ForkKV~\cite{wang2026forkkv}---compose multiple mechanisms under one control plane, a sign that the field is moving from point solutions toward integrated stacks. We note this as an observed trend and stop short of forecasting where it lands.

Closing the measurement gaps is the precondition for making the five archetypes useful as a guide for evaluating future systems and not only as a retrospective map. Until the community collects and publishes completion semantics (MG3), reuse-distance distributions (MG4), and KV-aware public traces (MG7), cross-archetype comparison will remain qualitative.

\bibliography{references}
\bibliographystyle{acm}

\end{document}